\documentclass[preprint,12pt]{elsarticle}
\usepackage{amsmath}

\usepackage{graphicx}

\usepackage{units}
\usepackage{hyperref}
\usepackage{algorithmic,algorithm}
\usepackage{color}
\usepackage{amssymb,amsmath}

\usepackage{moreverb}

\newcounter{bla}

\newcommand{\eq}[1]{Eq.~(\ref{#1})}
\newcommand{\fig}[1]{Fig.~{\ref{#1}}}
\newcommand{\tab}[1]{Tab.~{\ref{#1}}}
\renewcommand{\sec}[1]{Sec.~{\ref{#1}}}

\newcommand{\SU}[1]{\mathrm{SU}(#1)}
\newcommand{\U}[1]{\mathrm{U}(#1)}

\newcommand{\be}{\begin{equation}}
\newcommand{\ee}{\end{equation}}

\renewcommand{\Re}{\mathfrak{Re}\,}
\newcommand{\tr}[1]{\,\mathrm{tr}\left[#1\right]}
\newcommand{\bigO}{\mathcal{O}}
\newcommand{\e}{\mathrm{\,e}}
\newcommand{\tmin}{t_{\mathrm{min}}}
\newcommand{\tmax}{t_{\mathrm{max}}}
\newcommand{\1}{\mathrm{1\hspace{-0.4em}1}}
\newcommand{\bv}{\begin{pmatrix}}
\newcommand{\ev}{\end{pmatrix}}
\renewcommand{\vec}[1]{\boldsymbol{#1}}

\newenvironment{gather+}{%
  \equation\gathered
}{%
  \endgathered\endequation
}
\newenvironment{gather-}[1][]{%
  \subequations #1\gather
}{%
  \endgather\endsubequations
}

\newenvironment{align+}{%
  \equation\aligned
}{%
  \endaligned\endequation
}

\newenvironment{align-}[1][]{%
  \subequations #1\align
}{%
  \endalign\endsubequations
}

\journal{Computer Physics Communications}

\begin{document}

\begin{frontmatter}



\title{Coulomb, Landau and Maximally Abelian Gauge Fixing in Lattice QCD with Multi-GPUs}


\author[a]{Mario Schr\"ock\corref{email1}}
\author[b]{Hannes Vogt\corref{email2}}

\cortext[email1] {\textit{E-mail address:} mario.schroeck@uni-graz.at}
\cortext[email2] {\textit{E-mail address:} hannes.vogt@uni-tuebingen.de}

\address[a]{Institut f\"ur Physik, FB Theoretische Physik, Universit\"at
Graz, 8010 Graz, Austria}
\address[b]{Institut f\"ur Theoretische Physik, Auf der Morgenstelle 14, 72076 T\"ubingen, Germany}

\begin{abstract}
A lattice gauge theory framework for simulations on graphic processing units (GPUs)
using NVIDIA's CUDA is presented.
The code comprises template classes that take care of an optimal data pattern to ensure
coalesced reading from device memory to achieve maximum performance.
In this work we concentrate on applications for lattice gauge fixing in 3+1 dimensional
SU(3) lattice gauge field theories.
We employ the overrelaxation, stochastic relaxation and simulated annealing algorithms
which are perfectly suited to be accelerated by
highly parallel architectures like GPUs.
The applications support the Coulomb, Landau and maximally Abelian gauges.
Moreover, we explore the evolution of the numerical accuracy
of the SU(3) valued degrees of freedom over the runtime of the algorithms
in single (SP) and double precision (DP). 
Therefrom we draw conclusions on the reliability
of SP and DP simulations and suggest a mixed precision scheme
that performs the critical parts of the algorithm in full DP while 
retaining 80--90\% of the SP performance.
Finally, multi-GPUs are adopted to overcome the memory constraint of single GPUs.
A communicator class which hides the MPI 
data exchange at the boundaries of the lattice domains, via the low bandwidth
PCI-Bus, effectively behind calculations in the inner part of the domain is presented.
Linear scaling using 16 NVIDIA Tesla C2070 devices and a maximum performance of 3.5 Teraflops
on lattices of size down to $64^3\times 256$ is demonstrated.
\end{abstract}

\end{frontmatter}

%


\section{Introduction}\label{sec:introduction}
Quantum Chromodynamics (QCD) is nowadays, 40 years after its birth,
widely accepted as the correct theory of the strong nuclear force
which binds the protons and neutrons in the cores of atoms.
The guiding principle in the construction of QCD was the local gauge symmetry
which has led before to the very successful theory
of quantum electrodynamics (QED) 
that describes the interactions of electrons and light.
Local gauge symmetry is
the freedom to perform a transformation of the vector fields 
of the theory, independently at each point of space-time,
without changing the physics the theory describes.

Lattice QCD which lives on a discretized
space-time background opposed to the continuous world of the original theory,
offers a formulation of the gauge theory that is well suited to be 
simulated on a computer and hence can be used to test the theory 
against experiment. Furthermore, lattice simulations can help to gain insights in the highly
nontrivial, nonperturbative regime of the interactions between quarks and gluons which are the degrees of
freedom of QCD.

The gauge symmetry, given below in its discrete version, 
states that physical observables will remain unchanged if a local 
transformation of the form 
\begin{equation}
	g(x) U_\mu(x) g(x+\hat\mu)^\dagger
\end{equation}
is being carried out.
Here, the gauge fields or \emph{link variables} $U_\mu(x)$ 
as well as the gauge transformations $g(x)$
are elements of the underlying
gauge group which is $\SU{3}$ in the case of QCD.
The index $\mu=0,\hdots,3$ refers to the direction in four dimensional space-time 
and with $x+\hat\mu$
we denote the neighbor lattice site of $x$ in the $\mu$-direction.
The link variables of lattice QCD are connected to the algebra valued continuum 
gauge fields $A_\mu(x)$
via 
\begin{equation}
	U_\mu(x) = \e^{iagA_\mu(x)}.
\end{equation}

Whereas physical observables that can be measured in experiments
must be independent of the gauge, fixing the gauge, i.e., choosing
a particular gauge transformation $g(x)$ for all $x$, is essential
when, e.g., studying gauge dependent quantities like the fundamental two point functions
of the theory.

As a typical example of a gauge condition that may be enforced at all space-time points $x$, 
we consider the manifestly covariant Landau gauge
\be\label{eq:landaugauge}
	\partial_{\mu} A_\mu(x)=0,
\ee
here stated in the language of continuum field theories.
As we will discuss in the next Section, the continuum gauge condition \eq{eq:landaugauge}
translates to a large scale optimization problem in lattice QCD
with $\bigO(VN_c^2)$ degrees of freedom where $V=N_s^3\times N_t$ is the $3+1$ dimensional 
lattice volume.
Consequently, the process of fixing the gauge on the lattice demands
a major part of the whole simulation's computer time and the possible
acceleration by highly parallel hardware architectures like
graphic processing units (GPUs) will be of great practical use.

A more conceptual issue of gauge fixing is that the set of gauge transformations $g(x)$
that fulfill a desired gauge condition is far from being unique.
The set of gauge equivalent configurations of a given gauge field is called
the gauge orbit. The gauge fixing condition can be depicted as a hypersurface living
in the space of all gauge fields. Each of the multiple intersections of the gauge orbit 
with the gauge fixing hypersurface is called a Gribov copy.

Gribov copies play a crucial role in restoring the BRST symmetry on the lattice: 
fixing a gauge via the Faddeev--Popov procedure on the lattice for a compact group boils down 
to inserting the sum over signs of the corresponding Faddeev--Popov determinants evaluated 
at all the Gribov copies. Neuberger \cite{Neuberger:1986xz} showed that the sum for any covariant gauge turns out 
to be zero for any standard model gauge group, $\SU{N}$, and for compact $\U{1}$, making the 
expectation value of a gauge-fixed observable 0/0. The zero comes up because each Gribov 
copy comes in pairs with opposite sign of the Faddeev--Popov determinant. This in turn makes it
 impossible to construct a BRST symmetry on the lattice. This is called the Neuberger 0/0 problem. 
Following a topological interpretation of the Neuberger 0/0 problem, 
in Refs.~\cite{vonSmekal:2007ns,vonSmekal:2008es} 
a modified lattice Landau gauge was proposed which evaded the problem. 
There, because the Faddeev--Popov is shown to be strictly positive (semi-)definite, 
the cancellation is avoided. However, it is yet to be shown that the number of Gribov 
copies in the modified lattice Landau gauge is independent of the background gauge field. 
Interestingly, recently, a deep relation between lattice gauge fixing and lattice 
supersymmetry has been proposed in Ref.~\cite{Catterall:2011aa,Mehta:2011ud}: 
the partition functions of a class of supersymmetric Yang--Mills theories can be 
viewed as a gauge fixing partition function \`a la Faddeev--Popov and the ``Gribov copies'' 
are then nothing but the classical configurations of the theory.

A possible way out of the problem of the existence of Gribov copies is to
restrict the gauge fixing hypersurface to a region which the gauge orbit intersects
only once. An example thereof is the so-called Fundamental Modular Region \cite{Zwanziger:1998ez}
which contains only that intersection of the gauge orbit which corresponds to the global
optimum of the gauge condition.
Unfortunately, no algorithm is known which finds the global optimum of the gauge condition
within finite simulation time. \emph{Simulated annealing}, however, has been shown to 
highly favor optima closer to the global optimum \cite{kirkpatrick, Bali:1996dm} and moreover it can 
be shown that, in the limit of infinite time,
simulated annealing actually converges to the global maximum.

Recently, the problem of counting Gribov copies has gained a renewed interest. 
In Refs.~\cite{Mehta:2010pe}, an explicit formula of the number of Gribov copies 
for any number of lattice sites is analytically derived for lattice Landau gauge for the 
one-dimensional compact $\U{1}$ case. In Refs.~\cite{Mehta:2009zv,Hughes:2012hg}, 
a novel method based on Algebraic Geometry~\cite{Mehta:2011xs,Mehta:2011wj,Mehta:2012wk}, 
which can count all the Gribov copies, was proposed. 
Although the method has only been able 
to work for small lattices, it is the only known method which guarantees to find all Gribov 
copies and hence it can work as a benchmark for other methods.
One such alternative method is plain brute force, i.e., running a standard optimization
algorithm over and over again from different starting points on the gauge orbit
and collecting the results consecutively.
Clearly, a high performance lattice gauge fixing code is essential for this task and since here
one primarily focuses on small lattices, GPUs are favorable given the fact that CPU parallelization
techniques are very limited for lattices of small extent.

In this work we present a set of applications for lattice gauge fixing based on the family
of relaxation algorithms and simulated annealing. 
The applications are based on the CUDA accelerated Lattice--Graz--T{\"u}bingen 
code\footnote{Available for download at \href{http://www.culgt.com/}{www.cuLGT.com}}
that is written in CUDA C/C++ and makes heavy use 
of template classes in order to facilitate the extension to a broad variety of 
applications.
Besides the standard relaxation algorithm \cite{Mandula:1987rh}, 
we support overrelaxation \cite{Mandula:1990vs}
and stochastic relaxation \cite{deForcrand:1989im} to overcome the problem of critical slowing down.
Moreover, the simulated annealing algorithm \cite{kirkpatrick} with a heatbath kernel and microcanonical updates  which 
increases the probability to reach the Fundamental Modular Region
has been implemented and tested.
The code can be used to fix gauge configurations to the covariant
Landau gauge $\partial_{\mu} A_\mu=0$, $\mu=0,\hdots,3$, the Coulomb gauge  $\partial_i A_i=0$, $i=1,2,3$
and the maximally Abelian gauge.

Previous utilizations of GPUs in the field of lattice QCD mainly focused on solvers of linear systems
\cite{Clark:2009wm, Alexandru:2011ee}

A first attempt of porting lattice gauge fixing with the overrelaxation 
algorithm to the GPU has been reported in \cite{Schrock:2011hq, Schrock:2012rm}.
An alternative approach based on the 
steepest descent method has been presented in \cite{Cardoso:2012pv}.
For a more general discussion of lattice gauge fixing and its problems 
we refer the reader to \cite{Giusti:2001xf}.

The remainder of this work is organized as follows:
in \sec{sec:thealgo} the optimization problem is stated and the algorithms of
choice are presented.
In \sec{sec:cuda} we summarize some hardware properties of the NVIDIA GPUs that we 
use for our investigation and moreover briefly discuss NVIDIA's programming environment 
\emph{CUDA}.
Next, in \sec{sec:implementation}, we give details of our implementation and the cuLGT framework
and moreover discuss numerical accuracy issues. 
To overcome the memory constraint of single GPUs we extend our implementation
to support multi-GPUs; all details thereto are presented in \sec{sec:multiGPU}.
Finally, in \sec{sec:Results} we show various performance results for single and multiple GPUs
and furthermore present some convergence results of the algorithms.
In \sec{sec:summary} we summarize and conclude.

\section{The algorithms}\label{sec:thealgo}
In this Section we will first summarize the defining equations of the optimization
problem. Subsequently, we discuss the various flavors of the update kernels and
finally we list the main underlying algorithm of this work explicitly in 
terms of pseudo-code.

\subsection{The gauge functionals}
On the lattice, enforcing a gauge condition, e.g., \eq{eq:landaugauge}
is equivalent to maximizing the corresponding gauge functional.
We support three different kinds of gauge conditions and here we give the 
related gauge functionals and moreover a measure of
the iteratively achieved gauge quality that can serve as a stopping criterion
for the algorithm.

\subsubsection{Coulomb and Landau gauge}
The continuum Landau gauge condition, \eq{eq:landaugauge},
is fulfilled if and only if the lattice gauge functional
\begin{equation}\label{eq:landau_functional}
	F_{\mathrm{Landau}}^g[U] = \frac{1}{N_cN_dV}\Re\sum_{\mu, x} \tr{U^g_\mu(x)},
\end{equation}
resides in a stationary point
with respect to gauge transformations $g(x)\in\mathrm{SU}(N_c)$. 
In the above equation we made use of the short hand notation
\begin{equation}
 U^g_\mu(x) \equiv g(x) U_\mu(x) g(x+\hat\mu)^\dagger.
\end{equation}
Furthermore, with $N_c$ we denote the dimension of the gauge group $SU(N_c)$, $N_c=3$ for QCD,
$N_d$ is the number of space-time dimensions, ($N_d=4$ for our work) and
$V$ is the number of lattice points.
When switching to Coulomb gauge, all that changes is that the sum in \eq{eq:landau_functional}
becomes limited to the spatial components of the Dirac index $\mu$,
thus leaving out the temporal one.
Consequently, the optimization of \eq{eq:landau_functional} for 
Coulomb gauge can be performed independently on different time-slices.

A measure $\theta$ of how well the Landau/Coulomb gauge condition is satisfied
on a given gauge field configuration
is the average $L_2$-norm of the gauge fixing violation $\Delta(x)$, i.e.,
the discrete derivative of the continuum gauge fields
\be
	\Delta(x)\equiv \sum_\mu\left(A_\mu(x)-A_\mu(x-\hat \mu) \right) =0,
\ee
\begin{equation}
\theta\equiv \frac{1}{N_cV}\sum_{x}\tr{\Delta(x)
\Delta(x)^\dagger}.
\end{equation}

\subsubsection{Maximally Abelian gauge}
The gauge functional for the maximally Abelian gauge is, in the case of $\SU{2}$, given by
\begin{equation}\label{eq:MAG_SU2_functional}
	F_{\mathrm{MAG2}}^g[U]=\frac{1}{2N_dV}\sum_{x,\mu}\tr{\sigma_3 U_\mu(x) \sigma_3 U_\mu(x)^\dagger}
\end{equation}
where $\sigma_3$ is the diagonal matrix of the three Pauli matrices that correspond to the generators of $\SU{2}$.
Equivalently, in the case of $\SU{3}$ the gauge functional reads
\begin{equation}\label{eq:MAG_SU3_functional}
	F_{\mathrm{MAG3}}^g[U]=\frac{1}{3N_dV}\sum_{x,\mu}
	       \tr{ \lambda_3 U_\mu(x) \lambda_3 U_\mu(x)^\dagger }
	      +\tr{\lambda_8 U_\mu(x) \lambda_8 U_\mu(x)^\dagger }
\end{equation}
where $\lambda_3$ and $\lambda_8$ build the Cartan subalgebra of $\SU{3}$.
Maximizing \eq{eq:MAG_SU3_functional} is equivalent to 
minimizing the off-diagonal components $A_\mu^{(i)}(x),\;i\neq 3,8$ of the continuum gauge fields
\be
	A_\mu(x) = \frac{1}{2}\sum_{i=1}^8 \lambda_i A_\mu^{(i)}(x).
\ee
Note that maximizing \eq{eq:MAG_SU2_functional} or \eq{eq:MAG_SU3_functional}, respectively,
is equivalent to maximizing the squares of the diagonal of each gauge link
\begin{equation}\label{MAG_functional}
	F_{\mathrm{MAG}}^g[U]=\frac{1}{N_cN_dV}\sum_{x,\mu,i} \left|\left(U_\mu(x)\right)_{ii}\right|^2
\end{equation}
which is the gauge functional that we use in practice.

When the $\SU{2}$ gauge functional \eq{eq:MAG_SU2_functional} is stationary
with respect to gauge transformations, then the off-diagonal elements of
\be
	X(x) = \sum_\mu \left( U_\mu(x)\sigma_3 U_\mu(x)^\dagger
	      + U_\mu(x-\hat\mu)^\dagger \sigma_3 U_\mu(x-\hat\mu) \right)
\ee
must vanish \cite{Stack:2002sy}.
Thus, for $\SU{2}$ we can use 
\be
	\theta = \frac{1}{N_cV}\sum_{x} \left| (X(x))_{12} \right|^2
\ee
as a measure of the gauge quality. The off-diagonal element $(X(x))_{12}$
reads explicitly
\begin{gather+}
	(X(x))_{12} = \sum_\mu 2\big(u_{\mu,0}(x)u_{\mu,2}(x) + u_{\mu,1}(x)u_{\mu,3}(x) \\
	            - iu_{\mu,0}(x)u_{\mu,1}(x) - iu_{\mu,2}(x)u_{\mu,3}(x) \\
	            + u_{\mu,0} (x-\hat\mu) u_{\mu,2} (x-\hat\mu) + u_{\mu,1} (x-\hat\mu) u_{\mu,3} (x-\hat\mu) \\
	            + iu_{\mu,0} (x-\hat\mu) u_{\mu,1} (x-\hat\mu) -iu_{\mu,2} (x-\hat\mu) u_{\mu,3} (x-\hat\mu) \big)
\end{gather+}
where we adopted the Cayley--Klein parametrization
\be
	U_\mu = \bv u_{\mu,0}+iu_{\mu,3} && u_{\mu,2} + iu_{\mu,1} \\ -u_{\mu,2}+iu_{\mu,1} && u_{\mu,0}-iu_{\mu,3} \ev.
\ee
For $\SU{3}$, we use equivalently
\be
	\theta = \frac{1}{N_cV}\sum_{x} \left| (X(x))_{12} + (Y(x))_{12} + (Z(x))_{12} \right|^2
\ee
where the matrices $X(x), Y(x), Z(x)\in\mathrm{SU}(2)$ stem from the three $\SU{2}$ subgroups of $\SU{3}$.

\subsection{Relaxation}
Now that we stated the optimization problem, we can proceed with presenting
the algorithms which we will use to find a solution to the problem before
we will discuss the implementation with CUDA in the next Section.

The main idea of the relaxation algorithm is to sweep over the lattice site by site
while optimizing the gauge functional locally. Thereby, as we will see below, can all 
sites of one of the two \emph{parity subsets} (think of a checker board decomposition) 
be optimized at the same time
since the newly generated local optimum is a function of the nearest neighbors only.

In the following we will discuss the calculation of the local optimum
separately for Coulomb/Landau gauge and the maximally Abelian gauge.

\subsubsection{Coulomb and Landau gauge}
Instead of taking the complete global gauge functional into account,
\begin{equation}\label{eq:landau_loc}
	F_{\mathrm{Landau}}^g[U] = \frac{1}{2N_cN_dV} \Re\sum_{x} f_{\mathrm{Landau}}^g(x),
\end{equation}
the relaxation algorithm aims at optimizing the value of $F^g[U]$ locally, i.e., for all $x$ 
the maximum of 
\be
	f_{\mathrm{Landau}}^g(x) = \Re\tr{g(x)K(x)}
\ee
is sought. Here we introduced
\begin{equation}\label{Kx}
	K(x):= \sum_\mu\Big( U_\mu(x) g(x+\hat\mu)^\dagger 
		+ U_\mu(x-\hat\mu)^\dagger g(x-\hat\mu)^\dagger\Big)
\end{equation}
where the sum runs over all space-time indices for Landau gauge and
for Coulomb gauge it leaves out the temporal index.
The local maximum thereof is,
in the case of the gauge group $\mathrm{SU}(2)$,
simply given by
\begin{equation}\label{localsolution}  
g(x) = K(x)^\dagger/\sqrt{\det{K(x)^\dagger}}.
\end{equation}
For the gauge group $\mathrm{SU}(3)$ (QCD) one iteratively operates in the 
three $\mathrm{SU}(2)$ subgroups
\cite{CabibboMarinari1982} and thereby optimizes the local $\SU{3}$ gauge functional.

\subsubsection{Maximally Abelian gauge}
Similarly as for the Coulomb and Landau gauges, 
the goal is to maximize the gauge 
functional \eq{eq:MAG_SU2_functional} locally.
Again, we only need to know how to achieve this for $\SU{2}$
and then we can operate in the $\SU{2}$ subgroups of $\SU{3}$
for applications in QCD.

Thus, for a given site $x$ 
we want to maximize 
\begin{gather+}\label{MAG_SU2_functional_local}
	f_{\mathrm{MAG2}}^g(x) = \sum_\mu \mathrm{tr}\big[\sigma_3 g(x)U_\mu(x) \sigma_3 U_\mu(x)^\dagger g(x)^\dagger\\
		   + \sigma_3 U_\mu(x-\hat\mu)^\dagger g(x)\sigma_3 g(x)^\dagger U_\mu(x-\hat\mu)\big].
\end{gather+}

Let us focus on the part of \eq{MAG_SU2_functional_local} 
with the up-going links only, i.e., 
the first term in the sum;
the second term of \eq{MAG_SU2_functional_local} can be treated equivalently.

For the following discussion 
it will be useful to switch to the Cayley--Klein parametrization 
of $g(x)$ and $U_\mu(x)$,
\be
	g = g_0 \1 + i \sum_{i=1}^3 g_i\sigma_i 
		= \bv g_0+ig_3 && g_2 + ig_1 \\ -g_2+ig_1 && g_0-ig_3 \ev
\ee
and 
\be
	U_\mu = \bv u_{\mu,0}+iu_{\mu,3} && u_{\mu,2} + iu_{\mu,1} \\ -u_{\mu,2}+iu_{\mu,1} && u_{\mu,0}-iu_{\mu,3} \ev,
\ee
respectively, where for a simpler notation we suppressed the space-time argument $x$.

Taking the fact that transformations proportional to $\sigma_3$ leave the
functional \eq{eq:MAG_SU2_functional} unchanged into account (thus setting $g_3=0$)
one obtains
\begin{align+}\label{up}
	f^{\mathrm{up}}_{\mathrm{MAG2}}(x) = \sum_\mu&-2 \Big(4 g_0 (g_1 u_{\mu,0} u_{\mu,1}+g_2 u_{\mu,0} u_{\mu,2}-g_2 u_{\mu,1} u_{\mu,3}+g_1 u_{\mu,2} u_{\mu,3})\\
		&+g_0^2 \left(-u_{\mu,0}^2+u_{\mu,1}^2+u_{\mu,2}^2-u_{\mu,3}^2\right)\\
		&+\left(g_1^2+g_2^2\right) \left(u_{\mu,0}^2-u_{\mu,1}^2-u_{\mu,2}^2+u_{\mu,3}^2\right)\Big).
\end{align+}
Using a matrix/vector notation with $\vec g^T\equiv (g_0,g_1,g_2)^T$ the latter
can be written as
\be\label{upMatrix}
	f^{\mathrm{up}}_{\mathrm{MAG2}}(x)  = 2 \vec g^T 
	\bv D && E && F \\
	    E && -D && 0\\
	    F && 0 && -D\ev \vec g
\ee
where we defined
\begin{align}
	D &= \sum_\mu \left(u_{\mu,0}^2 + u_{\mu,3}^2 -\frac{1}{2}\right) \\
	E &= 2\sum_\mu\left(-u_{\mu,0}u_{\mu,1} - u_{\mu,2}u_{\mu,3}\right) \\
	F &= 2\sum_\mu\left(-u_{\mu,0}u_{\mu,2} + u_{\mu,1}u_{\mu,3}\right)
\end{align}
whereby in $D$ we used $\det{U_\mu} = u_{\mu,0}^2 + u_{\mu,1}^2 + u_{\mu,2}^2 + u_{\mu,3}^2=1$.

Then the maximum of \eq{up} is found when $\vec g$ is set to the eigenvector of the matrix
of \eq{upMatrix} corresponding to the largest eigenvalue.
The largest eigenvalue is $\lambda=\sqrt{D^2+E^2+F^2}$ and the corresponding 
eigenvector is
\be
	\left( D + \sqrt{D^2+E^2+F^2}\;,\; E\;,\; F \right)^T.
\ee

We refer the reader to \cite{Stack:2002sy}
for more practical details related to the maximally Abelian gauge.

\subsubsection{Overrelaxation}
In order to reduce the \emph{critical slowing down} of the relaxation algorithm on large
lattices, the authors of \cite{Mandula:1990vs} suggested to apply an overrelaxation algorithm which replaces
the local gauge transformation $g(x)$ by $g^\omega(x),\;\omega\in[1,2)$ 
in each step of the iteration.
In practice the exponentiation of the gauge transformation will be done to first order.

\subsubsection{Microcanonical steps}
Applying a gauge transformation $g^\omega(x)$ with $\omega=2$ leaves the
Landau/Coulomb gauge functional invariant but these so-called microcanonical steps have the beneficial property
to lead to a faster decorrelation and thus to faster convergence of the functional from which
the simulated annealing algorithm will profit.

\subsubsection{Stochastic relaxation}
The stochastic relaxation algorithm replaces the local gauge update $g(x)$
by a microcanonical step $g^2(x)$ with probability $p$ and
can lead to faster convergence on large lattices.

\subsection{Simulated annealing}\label{sec:SA}
Annealing is a method in condensed matter physics to bring certain materials
in their ground state by first heating them above their melting point and
subsequently cooling them down very slowly. It is crucial hereby that
the system is given enough time to thermalize at each temperature step.
If so, the atoms will arrange themselves in such a way that the macroscopic
system ends up in its -- or at least close to its -- lowest energy state.

The authors of \cite{kirkpatrick} developed an analogy of annealing
and mathematical optimization problems.
Following this analogy, the function which is to be optimized corresponds
to the energy of the solid and the optimum is the ground state.

The algorithm then simply performs local \emph{Metropolis} updates
where the acceptance probability of a random local gauge update $g(x)$
is given by
\be
	P[g(x)] =
	\begin{cases}
		\hfill 1 \hfill & \mathrm{ if }\quad  f^g(x) \geq f(x)\\
		\exp\left(\frac{f^g(x)  -   f(x)}{T}\right) 
		& \mathrm{ else. }
	\end{cases}
\ee
Thus, while in a hot temperature regime, the algorithm accepts a worsening of 
the local gauge functional with a non-vanishing probability which ensures
that the algorithm may overcome local extrema in order to increase
the probability to find
the global optimum.

In practice, the Metropolis update gets replaced by \emph{heatbath} updates
that generate the new gauge transformation directly with the right Boltzmann like
probability distribution.\footnote{We use the \emph{Philox} RNG of the \emph{Random123 library}\cite{Salmon:2011:PRN:2063384.2063405} to generate random numbers in the heatbath kernel.}
In order to reach quicker thermalization at each
temperature step, we perform three microcanonical steps after each change
in temperature. Note that simulated annealing will never reach the required
gauge precision $\theta$ very accurately, instead relaxation or
overrelaxation which can be regarded as simulated annealing in the limit
of zero temperature, should be run subsequently to fully reach the required 
precision. 
See also \sec{SATempDep}.

\subsection{Putting things together}\label{algo_sumup}
After we have listed the details of the underlying large scale optimization
problem and the techniques to perform local optimizations, we are now in 
the position to consider the global optimization algorithm.

As mentioned before,
due to the strict locality of the family of relaxation algorithms 
and the simulated annealing algorithm,
we can perform a checkerboard decomposition of the lattice and 
operate on all sites of one of the two sublattices 
\emph{even} and 
\emph{odd}\footnote{The sum over the space-time indices $t+x+y+z$ determines whether a site is considered 
\emph{even} or \emph{odd}.} concurrently.
All of the above mentioned algorithms have the same underlying structure which
is depicted in Alg. \ref{alg1}.

\begin{algorithm}
	\caption{}
	\label{alg1}
	\begin{algorithmic}                 
		\WHILE{precision $\theta$ not reached}
			\FOR{sublattice = even, odd} 
				\FORALL{$x$ of sublattice}
					\FORALL{$\SU{2}$ subgroups}
						\STATE local optimization: find $g(x)\in\SU{2}$\hfill {Step 1.}
						\STATE which is a function of $U_\mu(x),\;U_\mu(x-\hat\mu)$
						\FORALL{ $\mu$}
							\STATE apply $g(x)$ to $U_\mu(x),\;U_\mu(x-\hat\mu)$\hfill {Step 2.}
						\ENDFOR
					\ENDFOR
				\ENDFOR
			\ENDFOR
		\ENDWHILE
	\end{algorithmic}
\end{algorithm}

We want to stress that the difference of the various update algorithms as well as
the difference between the gauges under consideration
lies exclusively in
Step 1 whereas,
as we list explicitly in \ref{sec:countingFlops},
the main work of the algorithm lies in Step 2  which
is independent of the update type and of the target gauge.

\section{CUDA}\label{sec:cuda}
Here we briefly introduce the CUDA (Compute Unified Device Architecture) programming
model and summarize the hardware properties of the GPUs we adopt in our study.

\subsection{The programming model} \label{sec:cuda:model}
The CUDA model demands the division of the underlying problem into subproblems,  so-called \emph{thread blocks}, that can be treated independently from each other in parallel. These thread blocks, on the other hand, are ensembles of threads and the threads within a thread block may communicate with each other through shared memory. The independent thread blocks then form a so-called \emph{grid}. This model is very flexible and allows the user to run a CUDA application on different hardwares (meaning different number of streaming multiprocessors (SMs) and CUDA cores) without the need for major adjustments.\footnote{Since we exclusively adopt devices of the Fermi generation, the characteristic of the SMs is always the same for our tests, see \sec{sec:cuda:hardware}.}
This abstraction layer is introduced into the C language by defining a new set of functions which are called \emph{kernels} and are identified by the \emph{\_\_global\_\_} declaration specifier. The kernel is executed $N$ times where
\begin{equation}
	N = \text{block size} \times \text{grid size}
\end{equation}
and each kernel call is a thread in the nomenclature introduced above. For the invocation of the kernel a new syntax is introduced where the block size and the total number of blocks (grid size) is specified. 
A unique index is given to each thread to assign, e.g., different memory addresses to different threads.

A group of 32 threads (the number depends on the hardware generation) of the same block are tied together to what is called a \emph{warp}. The operations of all threads within a warp are executed simultaneously as long as they follow the same instruction path. Otherwise, the operations become serialized resulting in up to 32 cycles instead of one, a \emph{warp divergence} occurs.

To efficiently hide memory latencies it is inevitable to have many warps active at the same time on a SM. The possible number of active blocks (or warps) depends on the available hardware that has to be divided among the threads, e.g., it depends on how many registers and how much shared memory is needed for an individual kernel.

\subsection{Memory Layout}
In the CUDA terminology the CPU on which the CUDA application is run is called the \emph{host}, whereas the GPU is called \emph{device} and the associated memory is called host and device memory, respectively. Communication between host and device memory is the main bottleneck. Although, for many single GPU implementations, communication is only necessary in the beginning and in the end of an application. 
How one effectively can deal with communication from device to device through the host memory in 
multi-GPU simulations is discussed in \sec{sec:multiGPU}.
The part of device memory that is accessible from the host as well as from all CUDA threads is called \emph{global memory}. Global memory is allocated by a command in the host code. Each thread may then allocate its private \emph{local memory} which resides in the same physical memory as global memory. Global and local memory are both cached in a L1 and L2 cache by default (for Fermi), on a cache miss the latency to device memory is very high. For most applications the bandwidth to device memory is another limiting factor, although it is large compared to a common CPU to RAM bandwidth.

For communication within a block \emph{shared memory} can be used. Shared memory has a very low latency since it resides in the same hardware as the L1 cache.

\subsection{Hardware} \label{sec:cuda:hardware}
We adopt four different NVIDIA Fermi GPUs for our study, the GTX~480 and GTX~580 from the consumer section and moreover the Quadro~4000 and the Tesla~C2070 from the scientific/HPC section.
The Tesla~C2070, opposed to the consumer cards, supports
ECC (error correcting code) protection for DRAM.
Recently, the successor of the Fermi architecture has been released (Kepler).
In \tab{tab:fermi} we give the data which is common to all Fermi GPUs, the hardware details of the individual devices are summarized in \tab{tab:hardware:devices}.

\begin{table}
	\center
	\begin{tabular}{l|c}
		compute capability           		& 2.0				\\\hline
		cores / SM						& 32 per SM		\\\hline
		warp size							& 32 \\\hline
		L1 cache / SM                		& 16 KiB or 48 KiB	 \\\hline
		shared memory / SM           		& 16 KiB or 48 KiB	\\\hline
		32-bit registers / SM        		& 32768 (32Ki)		\\\hline
		max. registers / thread      		& 63				\\
	\end{tabular}
	\caption{Specifications of the Fermi architecture.}
	\label{tab:fermi}
\end{table}


\begin{table}
	\center
	\begin{tabular}{l|c|c|c|c}
					                 		& GTX 480 		& GTX 580 		& Quadro 4000 	& Tesla C2070 \\\hline
		graphics clock						& 700 MHz		& 772 MHz		& 475 MHz		& 575 MHz \\\hline
		SMs									& 15 			& 16			& 8				& 14	 \\\hline
		total CUDA cores	           		& 480			& 512			& 256			& 448	 \\\hline
		device memory   		& 1.5 GiB 		& 1.5 GiB		& 2 GiB			& 6 GiB	 \\\hline
		memory bandw.           		& 177.4 GB/s	& 192.4 GB/s	& 89.6 GB/s		& 144 GB/s	\\\hline
		\end{tabular}
	\caption{Hardware details of the Fermi devices that we adopt in this work.}
	\label{tab:hardware:devices}
\end{table}

\section{Implementation details}\label{sec:implementation}
\subsection{Code design}
The design goal of our code was the minimization of local memory usage.
One of the main limiting factors of performance is the number of registers 
that are available per thread:
on Fermi GPUs, the latter bound is 63 registers of 32-bit each. 
If more variables (on the assembly level) are needed per thread, 
the registers are ``spilled'' to local memory. 
Local memory, as mentioned earlier, uses the same hardware as global memory 
and thus has the same (high) latency and bandwidth bounds. 
Besides register spilling another source of local memory usage may slow down 
the execution of a kernel: 
registers are not addressable and therefore will
arrays generally be placed in local memory. 
In order to capacitate the compiler to place arrays in registers, the size of the arrays 
and all index variables that access elements need to be computable at compile 
time\footnote{The latter statement implies that, for example, all for loops have to be unrolled.}. 
Early versions of our code fulfilled this requirement by manually unrolling all loops 
and using C macros to access array elements. 
The present code, however, uses template parameters for the lattice dimensions 
and the dimension $N_c$ of the gauge group $\SU{N_c}$. 
As a consequence, unrolling can perfectly be done by the compiler. 
This code design offers a very flexible setup for further lattice applications.

\subsection{Reduce memory transfers}\label{sec:redmemtransfer}
In order to reduce memory transfers between global memory and the kernel a 12 parameter representation
of the $\SU{3}$ matrices 
has been suggested \cite{DeForcrand:1986af, Clark:2009wm}, i.e., only two rows of the matrix
are stored and loaded. 
If we denote the first and the second row of the matrix with vectors $\vec{u}$ and $\vec{v}$,
respectively, then the third row is given by $(\vec{u}\times \vec{v})^*$.
The extra numerical work to reconstruct the full matrix is hidden since our kernels are bound by memory transactions and not by floating point operations. This optimization reduces the number of bytes to load and store per site from 576 bytes to 384 in single precision.

\subsubsection{Memory pattern}\label{sec:implementation:memory_layout}
Due to the hardware design of NVIDIA GPUs one has to adopt special memory layouts 
to efficiently utilize the memory bus to global memory. 
The peculiarity of these devices is that memory transactions of threads of the same warp 
are coalesced if they reside in the same 128-byte aligned segment in global memory. 
Consequently, neighboring threads (i.e. neighboring sites) should access neighboring memory addresses
to achieve high memory throughput.
A natural memory layout where the gauge links ($\SU{3}$ matrices) are 
stored in one block in memory does not fulfill these requirements, hence the
index order of the gauge fields in memory has to be adapted.

The authors of \cite{Cardoso:2012pv}, e.g, use the native CUDA datatype \emph{Float4} 
and therefore distribute the 12 real numbers of a $\SU{3}$ element (in the 12 parameter representation, see \sec{sec:redmemtransfer}) 
to three \emph{Float4} arrays. In contrast,
we build on a more flexible way by employing one large float or double array, respectively, 
in combination with an access pattern class that hides the memory layout from the user. 
This strategy allows us to easily change the memory layout depending on the properties
of the underlying application.

Here we list explicitly the memory patterns that are in use in our gauge fixing
applications, whereby the slowest running index is listed first:
\begin{itemize}
	\item \emph{StandardPattern} (natural layout): $t, x, y, z, \mu, i, j, c$
	\item \emph{GpuPattern}:  $\mu, i, j, c, p, \left[t, x, y, z\right]_p$ 
	\item \emph{GpuPatternTimeslice}: $t, \mu, i, j, c, p, \left[x, y, z\right]_p$
	\item \emph{GpuPatternParityPriority}: $p, \mu, i, j, c, \left[t, x, y, z\right]_p$
	\item \emph{GpuPatternTimesliceParityPriority}: $t, p, \mu, i, j, c, \left[x, y, z\right]_p$
	\label{lst:list:access_patterns}
\end{itemize}

where $i, j \in \left\{0,1,2\right\}$ are the matrix indices, $c$ identifies real ($c=0$) and imaginary ($c=1$)
 part of the complex number, $\mu \in \left\{0, \dots, 3\right\}$ is the direction of the link, 
$t$ is the index in temporal direction and $x,y,z$ correspond to the spatial components.
The index $p\in\left\{0,1\right\}$ stands for parity (\emph{even} and \emph{odd}, respectively) 
and in those patterns where it is in use
the space-time indices are split into two groups
\be
	\left[t, x, y, z\right]_p := \left\{\left. t, x, y, z \right|  t + x + y + z\mod 2 = p \right\}
\ee
and equivalently for $\left[x, y, z\right]_p$. Parity splitting is necessary to achieve coalesced
access to global memory, since we operate on the parity even and odd sublattices 
separately (see Alg. \ref{alg1}).

The \emph{GpuPattern} is used in the single GPU implementations of Landau and maximally Abelian gauge. 
For Coulomb gauge we employ the \emph{GpuPatternTimeslice} for the global gauge field array 
and the \emph{GpuPattern} in kernels that operate on $1\times N_s^3$ sublattices, i.e., 
within a single time-slice. 
To reduce memory traffic between the nodes in the multi-GPU implementation we adopt 
the \emph{GpuPatternTimesliceParityPriority}.
This allows that only the active parity of the time-slices at the border can be transferred between nodes.
All applications use the \emph{StandardPattern} to read and write files with the natural ordering.

All patterns assume that the global array is allocated for full 18 parameter $\SU{3}$ links 
although the applications load and store only 12 parameters.

\subsubsection{Representation of the $\SU{3}$ link variables}
We define a template class \emph{SU3} with a template parameter 
that determines the storage type. For matrices that reside in the global memory array we offer a class
\emph{Link} with three parameters: 
(1) the pointer to the global memory array, 
(2) a lattice site given in terms of an object of type \emph{SiteIndex} and 
(3) the direction $\mu$. 
No memory is allocated for \emph{SU3$\langle$Link$\rangle$} variables. 
For local matrices we offer the class \emph{Matrix} which allocates local memory (or uses registers
when possible) for matrix elements. 
Functions for copying between \emph{SU3$\langle$Link$\rangle$} and \emph{SU3$\langle$Matrix$\rangle$} 
are implemented, as well as functions to load only the first two rows 
(12 parameter representation) of the matrices as well as a function to restore the third row.

\subsection{The eight-threads-per-site strategy}
Within every iteration of the gauge fixing algorithms each site update needs its adjacent links. 
These are read from global memory and after the update they have to be written back to global memory.
After having restored the third line, these eight $\mathrm{SU}(3)$ matrices per site equal 
$8\times 18$ reals = 144 reals and therewith exceed the register limit of
63 per thread what results in register spills to global memory and as a consequence 
negatively effects the bandwidth bound performance of the kernel.

With the purpose of  reducing register spills, we switch to a finer parallelization granularity: instead of assigning one
thread to one lattice site we now tie eight threads to a single lattice site, i.e., one thread for each of the
eight matrices that are involved in a site update. As a result, each thread needs only 18 registers to store 
the gauge link.

In order to avoid warp divergences the kernel is invoked with a thread block size of $8\times 32 = 256$. 
By doing so, each of the eight warps takes care of one neighbor type of the 32 sites and thus all threads within one warp follow the same instruction path.

The gauge transformation is then accumulated in shared memory. Since one operates on
the $\mathrm{SU}(2)$ subgroups of $\mathrm{SU}(3)$ and an $\mathrm{SU}(2)$ matrix can
conveniently be represented by four reals, this requires $4\times 32=128$ reals or 512 bytes (SP) or 1024 bytes (DP) per
thread block. To avoid race conditions on the shared array the accumulation is done using the \emph{atomic\_add} function in single precision and by explicit serialization using \emph{\_\_syncthreads()} in combination with \emph{if}-statements in double precision\footnote{\emph{atomic\_add} is not supported for datatype \emph{double}.}.

The benefit of this strategy is that, in single precision, no register spillings occur at all 
if no further constraints on the kernel are applied (see \sec{sec:optimizations}) and
for double precision, register spills are drastically reduced.
The drawback of the current implementation compared to a more conventional one-thread-per-site strategy 
is that the number of simultaneously computed sites per multiprocessors is
decreased. Nevertheless this strategy results in 
a clear overall performance gain \cite{Schrock:2012rm}.

\subsection{Optimizations}\label{sec:optimizations}
Besides the aforementioned algorithmic optimizations we further tuned our code by optimizing 
the CUDA settings.

First of all we set \emph{launch bounds} to individual kernels: 
by specifying the number of threads per block and a 
minimum of active blocks a bound on the maximal register usage is given. 
Without \emph{launch bounds} the compiler uses 45 registers in the overrelaxation kernel for Landau and Coulomb
gauges, resulting in a theoretical occupancy of $42\%$. 
By setting the register limit to 32
the theoretical occupancy is increased to $67\%$ on the cost of a small amount of register spilling 
(24 byte stack frame, 24 byte spill stores, 40 byte spill loads).\footnote{The given values for register usage and spilling are for CUDA Toolkit 5.0 compiled for compute capability 2.0. They vary between different between CUDA 4.x and 5.0 but the optimal launch bounds are found to be the same.} 
The same settings are applied to the other gauge fixing kernels.

Fermi devices have a L1 cache that physically shares the same 64 KiB hardware (per SM) with shared memory. The size of the L1 cache and shared memory can be set by the user for each kernel. Since we only need 512 Byte shared memory per block and a maximum of 4 blocks is possible, we only need a total of 2 KiB shared memory per SM. This allows us to set the kernel to a \emph{prefer L1 cache} configuration which means 16 KiB shared memory and 48 KiB of L1 cache. With this setting the register spilling introduced by the launch bounds is cached more efficiently.

By a global compiler switch, the use of L1 cache can be set to either caching (default) or non-caching (-Xptxas -dlcm=cg) loads. By using non-caching loads our applications shows a small improvement in performance. This is due to the fact that the use of global memory is designed such that only in the beginning of each kernel the matrices are loaded to local memory (i.e., into registers) and after all operations are finished they are written back. In between there is no reuse of cached data and thus there is no benefit in caching at all. With non-caching load the L1 cache is solely used for register spilling and write-backs of register spills to the device memory are reduced or totally removed.

In all applications we compile with the \emph{use\_fast\_math} switch. Single precision operations are then replaced by faster implementations on the expense of precision though we did not experience any effects by this setting.  For double precision operations there is no such option.

\subsection{Numerical accuracy}
In the following we investigate the accumulation of numerical rounding errors within our lattice
gauge fixing applications. 
A suitable measure is the conservation of unitarity of the $\SU{3}$ matrices during the progress
of the algorithm through many iterations.
\begin{figure}[htb]
	\center
	\includegraphics[width=0.75\textwidth]{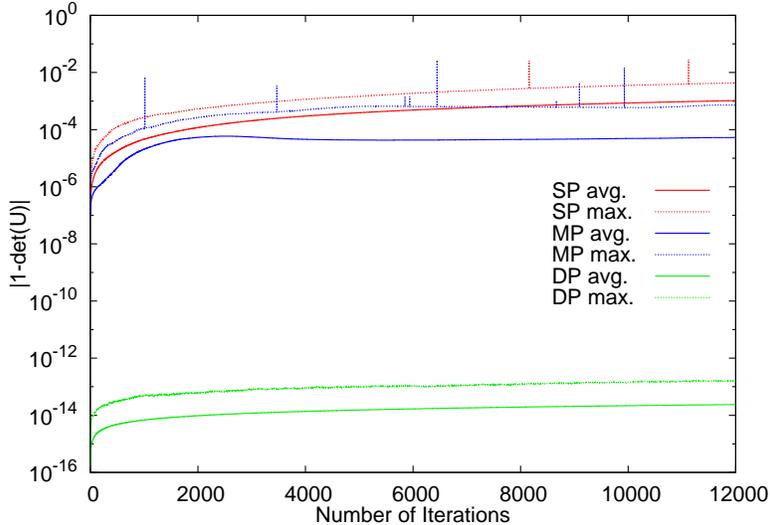}
	\caption{Conservation of unitarity ($|1-\det(U)|$) in SP, MP and DP}
	\label{fig:conserving_unitarity}
\end{figure}
In \fig{fig:conserving_unitarity} we show 
\be
	\frac{1}{4V}\sum_{\mu,x}\left|1-\det(U_\mu(x))\right| \quad\mathrm{and}\quad
	\max_{\mu,x}\left|1-\det(U_\mu(x))\right|
\ee
from a run over $12000$ iterations of the overrelaxation update on a $32^4$ lattice in single (SP)
and double (DP) floating point precision.
Moreover the plot shows lines corresponding to a mixed precision (MP) ansatz which calculates
the overrelaxation gauge update on the SP gauge fields in full DP 
(see~\fig{fig:calculateUpdate})
while the less precision demanding application of the gauge transformation to the 
links (Step 2 in Alg.~\ref{alg1}) is performed in SP.

In DP, both, the average and even the maximal value stay well below $10^{-12}$ whereas in SP
the error accumulates to the order $10^{-3}$. 
To overcome the loss of unitarity, one may use the unitarity as a constraint and thus 
reproject the links to $\SU{3}$ after a given number of iteration steps.

The peaks in the SP maximum lines are individual outliers that occur approximately every 1000 iterations
in one of the links of a $32^4$ lattice on our GTX~580, whereas they could not be detected on the Quadro~4000.

\begin{figure}[htb]
	\center
	\includegraphics[width=0.75\textwidth]{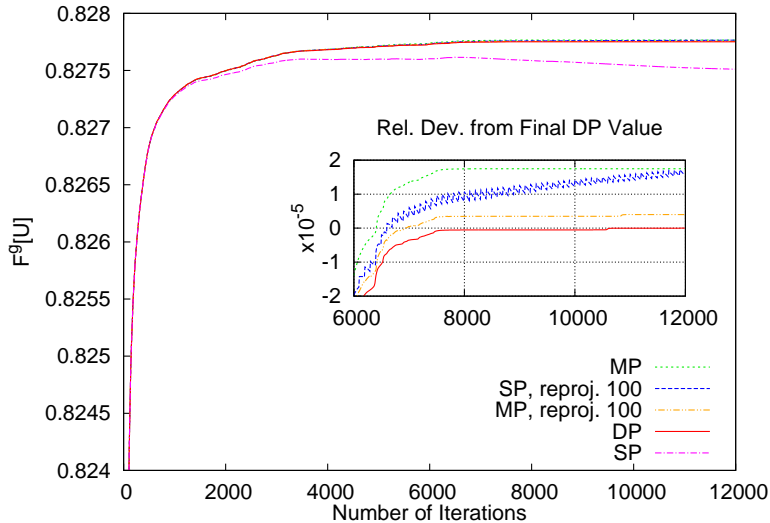}
	\caption{The value of the Landau gauge functional $F_{\mathrm{Landau}}^g[U]$
			as a function of the number of iterations of the overrelaxation kernel in
			single (SP), mixed (MP) and double precision (DP). 
			In addition, the evolution of the functional value is shown when a reprojection 
			to $\SU{3}$ is done every 100 iteration steps in SP and MP.
			The inner plot gives relative deviation of all curves (except SP without 
			reprojection) from the final functional value in DP.
		}
	\label{fig:preccomparison}
\end{figure}

Whether or not the loss of high precision unitarity in SP is of significance, depends
of course on the individual problem the code is applied to.
In \fig{fig:preccomparison} we show the value of the Landau gauge functional which is the
sensitive quantity in our applications,  in different precisions, again over 12000 
iterations\footnote{The gauge precision thereafter was $\theta< 6.0\times 10^{-11}$ for the run in DP.}
on a $32^4$ lattice.
It becomes obvious that SP without reprojection is not a good choice for lattice gauge fixing
since the value of $F^g$ even starts to decrease after around 3000 iterations.
The DP functional value increases monotonically and finally reaches a plateau, this fact together
with the previous mentioned maintenance of high precision unitarity lets us conclude that a
DP simulation, even without reprojection, is very accurate.
Thus, we can use the DP value as a benchmark for the other approaches.
In the inner plot of \fig{fig:preccomparison} we show the relative deviation of each curve to
the final DP result: SP with reprojecting to unitarity after every 100 steps and MP without
reprojection stay within a relative deviation of $2\times 10^{-5}$ and MP with reprojection
even within $5\times 10^{-6}$. Moreover, the MP line shares the same qualitative behavior as
the DP curve (monotonicity, convergence to a constant).

Therefore, our conclusion for the required floating point precision in lattice gauge fixing
is as follows: in case one is primarily interested to actually fix the gauge of a gauge field
configuration without being interested in the precise value of the resulting gauge functional,
SP with reprojection is fine. If it is required to obtain the gauge functional value within a precision of no more than 
$10^{-5}$, MP with reprojection is recommended since it retains most of 
the SP performance, as we will show in \sec{sec:Results}, opposed to DP which
should only be chosen when one depends on a high precision result in the value of the gauge functional.

\section{Multi-GPU}\label{sec:multiGPU}
In the following discussion we will replace the space-time argument $x=(\vec x, t)$
by the time argument $t$ alone wherever the $\vec x$ dependence is
of no significance in the given context.
Moreover, we will assume that one MPI process is assigned to one GPU device
and thus use the terms process and device interchangeably.

In order to share the work which has to be performed locally on each lattice site,
between several processes,
we adopt a straightforward domain decomposition:
we split the lattice of size $N_s^3\times N_t$ into $N_t/nprocs$ partitions,
where $nprocs$ denotes the number of processes involved in the parallelization.
For Coulomb gauge this splitting is trivial since, as we discussed above,
we can operate on the different time-slices separately and only need to
apply the final gauge transformation $g(t)$ of the time-slice $U_\mu(t)$
to the temporal components of the preceding time-slice $U_0(t-1)$.
This makes \emph{on-the-fly} communication between devices for Coulomb gauge fixing unnecessary.

Manifestly covariant gauges like the Landau gauge and the
maximally Abelian gauge, on the other hand, are more subtle.
Here, all four neighboring links
in the negative $\mu$-direction have to be collected on each site $x$ in order
to calculate the gauge update $g(x)$ which subsequently is applied
to all the eight links connected to the site $x$.
Thus, with the ansatz of splitting the lattice across the temporal direction,
we have to exchange the temporal components $U_0(x)$ of the gauge fields on
time-slices that lie at the boundary of two processes.

\subsection{Data exchange between neighboring devices}\label{sec:data_exchange}

If we label the minimum time-slice that resides on a given device with $\tmin$
and the maximum time-slice with $\tmax$,
then only the calculation of the local gauge transformations $g(\tmin)$ 
depends on the data exchange between different processes
since for its calculation the gauge links $U_\mu(\tmin-1)$ that reside on 
the neighbor process are needed.
Note that since we operate on the parity even and odd lattice sites consecutively,
the currently active parity of the time-slice $U_\mu(\tmax)$ is completely
unaffected by the exchange with the neighboring process that only touches the
passive parity part of $U_\mu(\tmax)$.
That means, on a given process, all time-slices except $\tmin$ can be updated 
without exchanging any information with the neighbor processes.
In order to update the $U_\mu(\tmin)$ on all devices, however, 
the following set of instructions has to be carried out on each device
in order to transfer the links $U_0(\tmax)$ of device $i$ to device $i+1$:
\begin{enumerate}
\item \emph{cudaMemcpyDeviceToHost} of $U_0(\tmax)$ (inactive parity)
\item \emph{MPI\_Send} of $U_0(\tmax)$ to device $i+1$ and
	\emph{MPI\_Recv} of $U_0(\tmin-1)$ from device $i-1$
\item \emph{cudaMemcpyHostToDevice} of $U_0(\tmin-1)$
\item update $U_\mu(\tmin)$ (active parity) which affects $U_0(\tmin-1)$ (inactive)
\item \emph{cudaMemcpyDeviceToHost} of $U_0(\tmin-1)$ (inactive parity)
\item \emph{MPI\_Send} of $U_0(\tmin-1)$ to device $i$ and
	\emph{MPI\_Recv} of $U_0(\tmax)$ from device $i+1$
\item \emph{cudaMemcpyHostToDevice} of $U_0(\tmax)$ 
\end{enumerate}

\subsection{Data pattern}
The memory pattern \emph{GpuPatternTimesliceParityPriority},
introduced in \sec{sec:implementation:memory_layout}, 
will be the pattern of choice for
applications that get accelerated by a time-slice splitted multi-GPU approach.
Not only is the time-index running slowest and thus allows
to handle different time-slices separately in the latter mentioned pattern,
moreover the time-slice
internal pattern is very advantageous: each time-slice is split into
its two parity parts of which each has the Dirac index $\mu$ running
fastest, followed by the row index of the individual gauge matrices.

This layout ensures that the data which has to be exchanged,
the first two rows (12 parameter representation) of the link variables
$U_0(\tmin)$ of a given parity, lie contiguous in device memory.
The size of the data block that has to be exchanged 
is then given by the size of a time-slice multiplied by
$1/2$ (parity), $1/4$ (Dirac index) and $2/3$ (12 parameter representation),
thus $1/12$ in total.

\subsection{Asynchronous memory transfers}
We target at hiding the data exchange between different devices
by overlapping them with calculations on the unaffected time-slices.
Replacing the CUDA function \emph{cudaMemcpy} with \emph{cudaMemcpyAsync}
results in a non blocking copying process from host to device or vice versa.
Making use of different \emph{cudaStream}s a device can then perform a
copying request and execute a kernel at the same time.

In order to investigate how many time-slices are needed per device
to fully hide the data exchange between two devices, we measured
the time for the execution of the overrelaxation kernel on one time-slice
and the time for a transfer of $1/12$ of a time-slice for different 
spatial lattice sizes $N_s^3$ and averaged the result over 1000 iterations,
see \tab{tab:data_exchange}.
\begin{table}
	\center
	\begin{tabular}{c|c|c|c|c|c}
	$N_s^3$ & D2H [$\mu s$] & H2D [$\mu s$] & kernel [$\mu s$] &
	 D2H/kernel & H2D/kernel \\\hline
		16  &  0.0398  & 0.0368 & 0.0209 & 1.90 & 1.76 \\
		32  &  0.2543  & 0.2276 & 0.1443 & 1.76 & 1.58  \\
		64  &  1.2510  & 1.1830 & 1.0489 & 1.19 & 1.13  \\
		128 &  8.9597  & 8.7169 & 8.3041 & 1.08 & 1.05
	\end{tabular}
	\caption{The time needed to copy the relevant part ($1/12$) of
	a time-slice from device to host (D2H) and host to device (H2D)
	compared to the time needed to update one time-slice with the 
	overrelaxation kernel (all in $\mu s$) averaged over 1000 iterations
	for different spatial volumes $N_s^3$. The two most right columns
	give the ratios.
	}
	\label{tab:data_exchange}
\end{table}
As we can read off from the table, the asynchronous kernel execution
on two time-slices takes longer then a device to host or host to device
copy process, respectively.  As discussed above, the necessary
data exchange between two devices includes in total four such copy processes
and thus eight time-slices are enough to reach a complete overlap of data
exchange from device to host (host to device) and calculations in the inner
part of the domain.

So far we neglected the data exchange via MPI between the two neighboring host processes.
As for the data exchange between host and device, here again it is advantageous
to use non blocking functions for the data exchange, i.e.,
\emph{MPI\_Isend} and \emph{MPI\_Irecv}. By doing so we can again overlap the data exchange
between the processes by calculations on time-slices that are not involved in the 
exchange.

In practice, we implemented the overlap of calculations with the data exchange between
the processes and between host and device as a method of a communicator class.
Then we only have to set up a certain update type (overrelaxation, simulated annealing etc.)
and the \emph{apply} method of the communicator object applies that update including full
overlap with the data exchange.

\section{Results} \label{sec:Results}
In this Section, we firstly examine the performance of the code on various devices
including multiple GPUs.
There, we pick the Landau gauge overrelaxation kernel as a representative for all kernels
and gauges.
Secondly, we outline a few sample results obtained
by the application of our lattice gauge fixing code.

\subsection{Performance on single GPUs} \label{sec:performance}
\begin{figure}[htb]
	\center
	\includegraphics[width=0.75\textwidth]{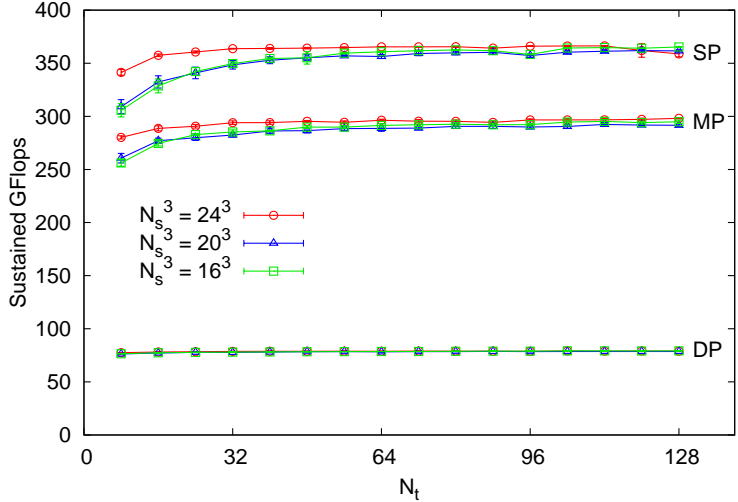}
	\caption{
		Performance of different spatial volumes as a function of 
		the temporal lattice extend in SP, MP and DP on a GTX~580.
		}
	\label{fig:performance_volume}
\end{figure}

In \fig{fig:performance_volume} we show the performance of the overrelaxation kernel on the GTX~580 for different spatial volumes as a function of the temporal lattice extent. 
The data stems from an average of one hundred repeated applications with 1000 iterations each.
We achieve up to 370 GFlops in SP, up to 300 GFlops in MP and 80 GFlops in DP. 
The maximum performance of 370 GFlops corresponds to an execution time of $\unit[6.4]{s}$
with the given lattice size and number of iterations.
For the smaller lattices the theoretical occupancy of the device is not reached and therefore the maximum performance is not achieved.
Apart from that, we find almost constant performance for all lattice volumes.

\begin{figure}[htb]
	\center
	\includegraphics[width=0.9\textwidth]{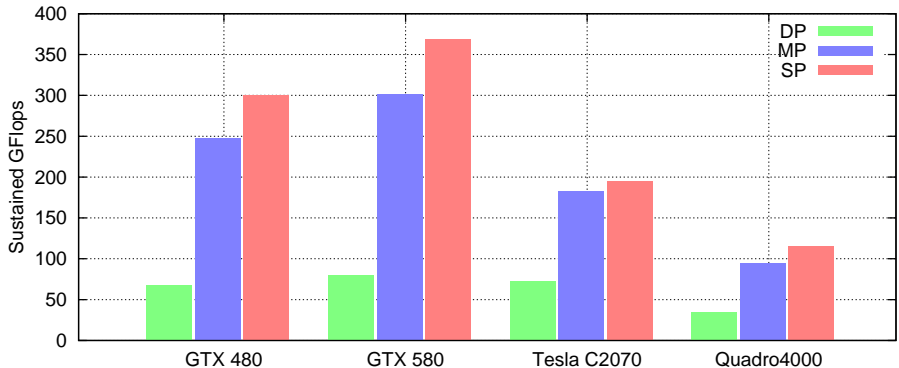}
	\caption{
			Performance of the Landau overrelaxation kernel on different NVIDIA devices
			in single (SP), mixed (MP) and double precision (DP) on a $32^4$ lattice.
		}
	\label{fig:performance_devices}
\end{figure}

In \fig{fig:performance_devices} we compare the performance of different Fermi devices
on lattices of size $32^4$. Our top performers in SP are the GTX~580 with nearly 370~GFlops, followed by the GTX~480
at around 300~GFlops. The difference between these devices results from the reduction in chip clock, number of SMs and bandwidth. 
The scientific GPUs are designed for a longer runtime and therefore the chip clock is remarkably lower. Thus, the performance of the
C2070 is only close to 200~GFlops, the Quadro~4000 is at around 120~GFlops. Noteworthy is the difference between single and double
precision: the theoretical ratio of SP to DP arithmetic operations for the scientific devices is 1:2, whereas the
consumer GPUs have a ratio of 1:8. Accordingly, the performance ranking changes: still the GTX~580 performs best with approximately
80~GFlops, now followed by the C2070, slightly faster than the GTX~480 at around 70~GFlops. Thus, even for the scientific GPUs the 
theoretical factor of a half compared to SP could not be reached. The reason is that approximately twice as many 
registers are needed in DP and therefore even for the maximum of 63 registers \emph{spilling} occurs. Additionally, the theoretical 
occupancy is reduced by the increase in registers.

The performance data given above is intended for comparing the algorithm on different architectures. Is is based on counting Flops as
described in \ref{sec:countingFlops}. The actual number of operations differs since we did not count the overhead for computing the 
third line reconstruction and we did not account for fused multiply-add operations. A true measure for the performance of our code
is the number of instructions per cycle (IPC). For the top performer, the GTX~580, the IPC in SP is $1.49$ which means roughly $75 \%$ of the peak performance,
since a maximum of 2 instructions is issued per cycle. On the other hand, the global memory throughput is 120 GB/s which is only approximately $60 \%$ of peak.
Combining these results, the most likely performance bound is not memory bandwidth, but \emph{memory latency} which could theoretically be cured by increasing 
occupancy. In practice this is not possible, since this would mean further decreasing registers per thread and thus introducing additional register spilling.

\subsection{Performance on multi-GPUs} \label{sec:MuliGPUperformance}
Our multi-GPU performance tests have been carried out on the ``mephisto'' cluster
at the University of Graz.
The cluster provides five compute nodes with four NVIDIA Tesla C2070 GPUs
and CUDA 5.0.
Moreover, each node offers two Intel Xeon Six-Core CPUs X5650 (``Westmere'') @ 2.67GHz on each node.
The nodes are connected via InfiniBand and OpenMPI 1.4.3 and CUDA 5.0 is installed.

\begin{figure}[htb]
	\center
	\includegraphics[width=0.75\textwidth]{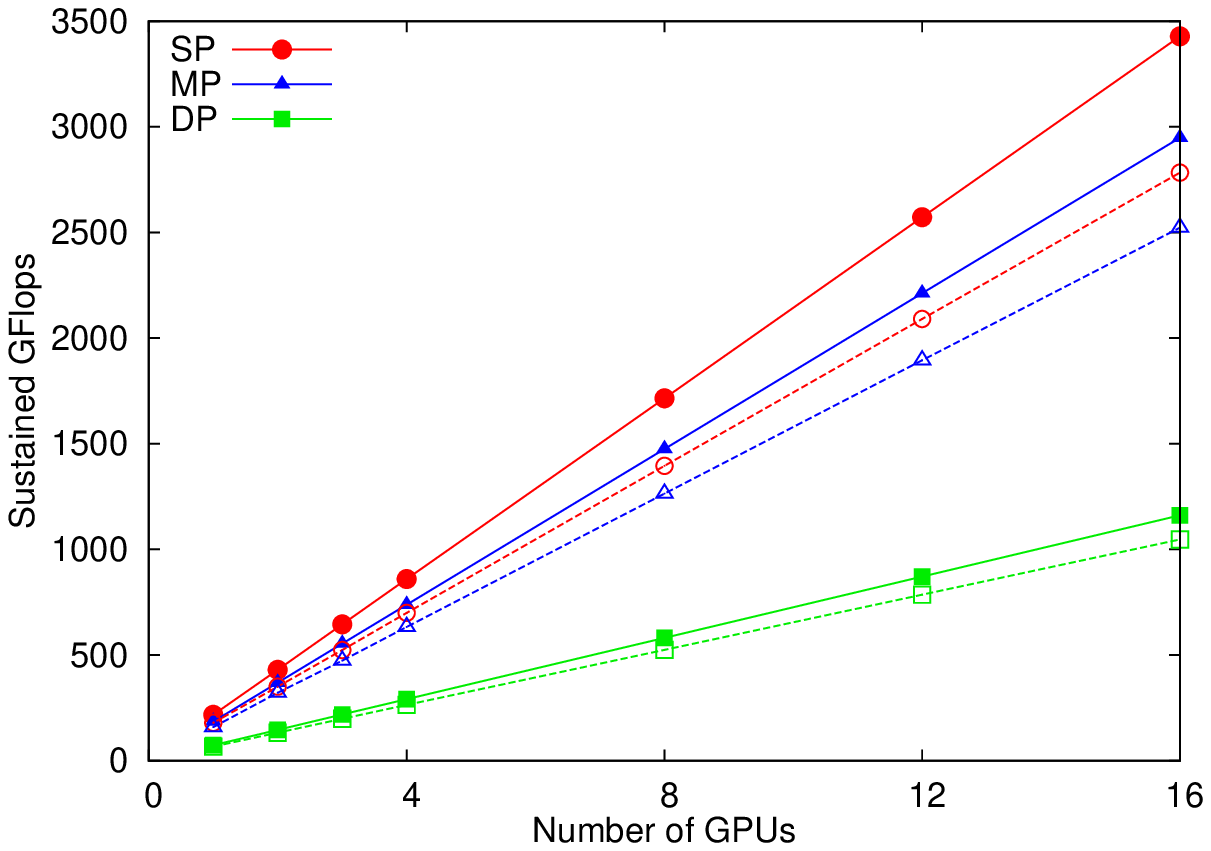}
	\caption{Weak scaling on the Tesla C2070 in single (SP), mixed (MP)
		and double precision (DP).
		The full symbols correspond to a lattice size of $64^3\times32$ per GPU and the
		open symbols to $48^4$ per GPU.
		}
	\label{fig:weakscaling}
\end{figure}

In the plot of \fig {fig:weakscaling} 
we show that linear weak scaling is reached
with this strategy. The test have been performed on lattices of size  $64^3\times32$ per GPU 
($64^3\times512$ in total with 16 GPUs) and $48^4$ per GPU ($48^3\times768$ 
in total with 16 GPUs). The higher performance of the spatial volume of $64^3$
is simply due to higher occupancy: since we operate on single time-slices 
at a time, the lattice of spatial size $48^3$ is not sufficient to efficiently 
occupy the device.

\begin{figure}[htb]
	\center
	\includegraphics[width=0.75\textwidth]{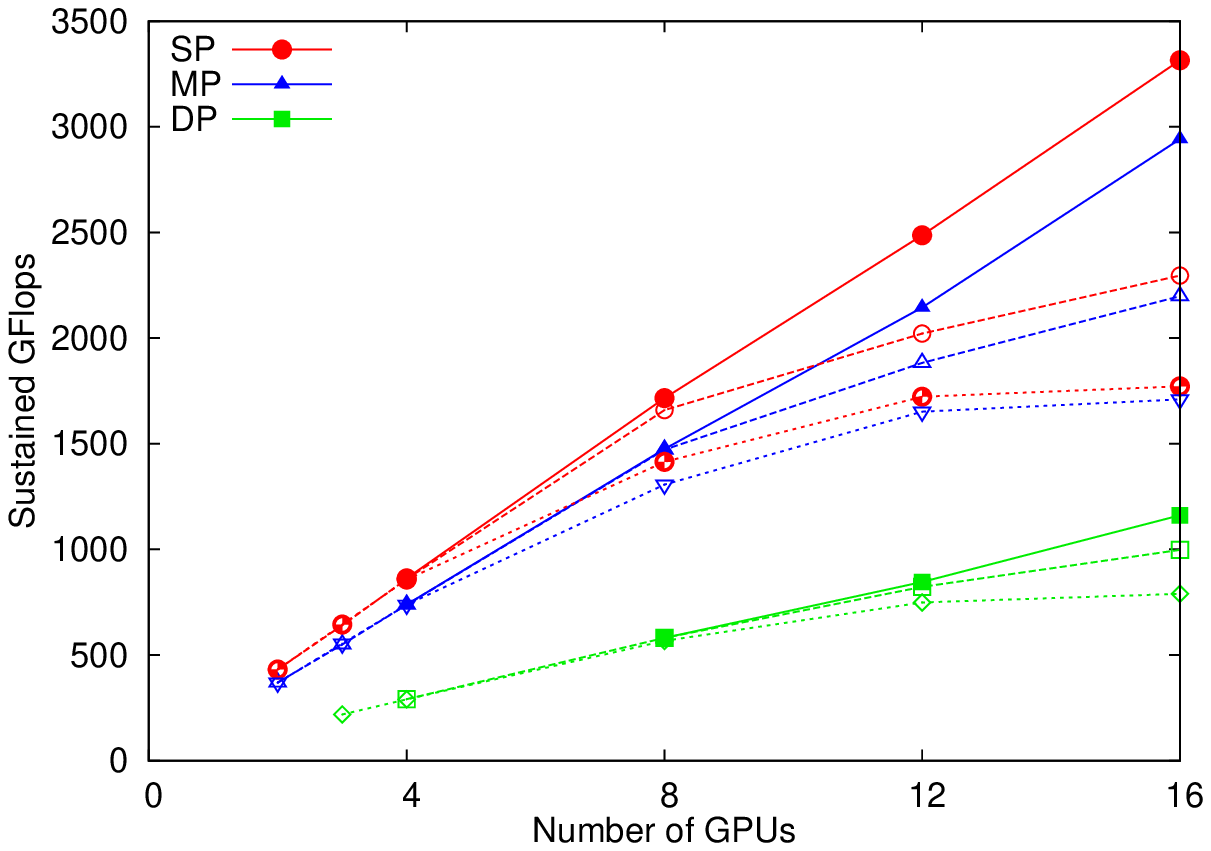}
	\caption{
		Strong scaling on the Tesla C2070. 
		The spatial lattice volume is kept fixed at $64^3$ and the total temporal extent
		varies for the three lines (per precision) from the top downwards
		$N_t=256, 128, 96$.
		}
	\label{fig:strongscaling}
\end{figure}

In \fig{fig:strongscaling}
strong scaling is tested. For a total lattice size of $64^3\times256$
we find close to linear strong scaling up to 16 GPUs which corresponds
to 16 time-slices per device. On a lattice of size $64^3\times128$
we find for 16 GPUs (eight time-slices per device) a performance loss
of 15--30\% (DP vs. SP).
When moving on to a smaller temporal lattice extent, $N_t=96$, the performance
decreases further. Moreover, for this lattice size, no gain in performance
is apparent when adopting 16 instead of 12 devices.

\subsection{Comparison to existing CPU code}
Lastly, we compare our performance to the overrelaxation kernel of the
\emph{FermiQCD} library \cite{FermiQCD}.
The \emph{FermiQCD} toolbox is open source (C++) and has been designed to be easy to use
while at the same time offering the user many applications for lattice QCD, in some applications 
at the expense of performance.
To our knowledge, it is the only publicly available code that supports
lattice gauge fixing with the overrelaxation algorithm in Landau 
gauge. We would be happy to compare our code to a wider range of implementations.

As test bed we chose an
Intel Xeon  Westmere CPU on the mephisto cluster, see \sec{sec:MuliGPUperformance}.
We run the \emph{FermiQCD} Landau gauge overrelaxation kernel in SP on a
lattice of size $32^4$ on a single core in avoidance to reflect parallelization artifacts.
Then we compare the performance to our code (same lattice size and precision)
from the Tesla C2070 that the cluster offers.

\emph{FermiQCD} reaches a performance of 0.414 GFlops and our code reaches
for this setup 195.08 GFlops. Thus, our
implementation executed on the Tesla GPU is equivalent to \emph{FermiQCD}
executed on $\approx 470$ CPU cores of the given type, under the naive assumption of linear
scaling for the CPU code.

\subsection{Temperature dependence of the simulated annealing algorithm}\label{SATempDep}
In \sec{sec:SA} we discussed the importance of keeping the temperature gradient small
in the simulated annealing.
Therefore, it is crucial to set up the right temperature interval in order not to 
waste many iteration steps in a temperature region where the gauge functional is insensitive to.
\begin{figure}[htb]
	\center
	\includegraphics[width=0.75\textwidth]{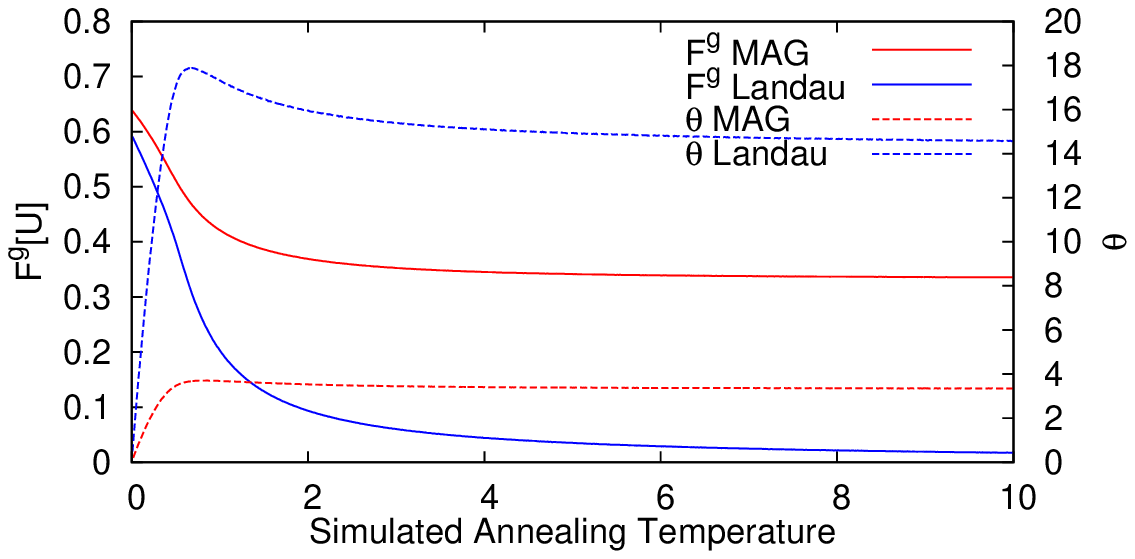}
	\caption{
		The temperature dependence in simulated annealing 
		of the gauge functional $F^g[U]$ and the gauge precision $\theta$
		of the Landau gauge and the maximally Abelian gauge.
		}
	\label{fig:SAtemp}
\end{figure}

In \fig{fig:SAtemp} we show an example of the evolution of the gauge functional $F^g[U]$ 
and the gauge precision $\theta$ of the Landau gauge and the maximally Abelian gauge.
The simulation has been performed on a \emph{hot gauge field}, i.e., having all gauge
links set to random $\SU{3}$ matrices. The lattice size is $32^4$ and for both cases
10,000 simulated annealing steps have been carried out.

As one can read of from the plot, in this case, the sensitive region where the gauge functional
changes most is for Landau gauge below $T< 4$ and for the maximally Abelian gauge
slightly lower,  $T< 2$.

\subsection{Cooling down to maximally Abelian gauge}
Here, we aim at reducing the time and number of iterations to
gauge fix a configuration to the maximally Abelian gauge.
We test overrelaxation versus a combination of simulated annealing, stochastic relaxation
and overrelaxation in terms of required number of iterations to gauge fix 
a sample gauge configuration with inverse coupling $\beta=5.7$ and lattice size $32^4$.

Both approaches use an overrelaxation parameter of $\omega=1.35$, the second
method starts off by applying 2000 simulated annealing steps including three
microcanonical updates after each step (i.e., 8000 steps in total).
Subsequently, a maximum of 2000 stochastic relaxation steps are applied
and lastly overrelaxation until the precision $\theta < 10^{-12}$ is reached. 
Method one directly applies the overrelaxation kernel until convergence.

We started both variants on 100 randomly chosen points on the gauge orbit.
Method one succeeded to find an optimum for 84 out of the 100 copies,
the remaining 16 got stuck at a value of $\theta \approx 10^{-7}$
until the algorithm was stopped after one hundred thousand iterations.
Method two was successful for 97 copies. 

\begin{table}
	\center
	\begin{tabular}{c|c|c}
	                            & OR & SA/SR/OR \\\hline\hline
	\# of converged copies      & 83 & 97 \\\hline
	\# of iterations & $272340\pm8405$ & $16701\pm2562$ \\\hline
	$F^g[U]$       &  $0.74356431(39697)$ & $0.74423815(10996)$ 
	\end{tabular}
	\caption{
		Comparing the application of the overrelaxation algorithm (OR) solely, to the subsequent
		application of simulated annealing (SA) with microcanonical steps, stochastic relaxation (SR)
		and OR on 100 copies of a gauge field of lattice size $32^4$.
	}
	\label{tab:coolMAG}
\end{table}

The average number of required iterations (the combined number of all updates)
is given in \tab{tab:coolMAG}, together with the final value of the gauge functional
$F^g[U]$.

As it is evidence from the data, the combined approach of simulated annealing, stochastic relaxation
and overrelaxation outperforms the pure overrelaxation method in terms of number of iterations
by a factor of almost two and moreover reaches an higher average value of the gauge functional
while bringing more gauge copies to converge.
The average time spend by the device (GTX~580) per gauge copy was four minutes 
for method two and slightly
below seven minutes for method one. It has to be stressed, however, that 
not all gauge copies converged
and hence these copies enter the average of the execution time with a biased weight since the kernel
was executed until the maximum number of iterations was reached.

\subsection{Towards the global maximum of the Landau gauge functional}
We take the same gauge field configuration with $\beta=5.7$ and lattice size $32^4$
of the previous subsection and now aim at finding Landau gauge Gribov copies with
gauge functional values as high as possible.
Three runs with 100 random starts
on the gauge orbit have been performed.
The difference of the three runs lies in the number of simulated annealing steps 
that are applied before the overrelaxation kernel takes over.
We apply zero, three thousand or ten thousand simulated annealing steps, respectively.
The temperature has been decreased from $4$ down to $10^{-4}$.
Each simulated annealing step is followed by three microcanonical updates.
Subsequently, we apply the overrelaxation kernel until $\theta < 10^{-10}$.

\begin{figure}[htb]
	\center
	\includegraphics[width=0.32\textwidth]{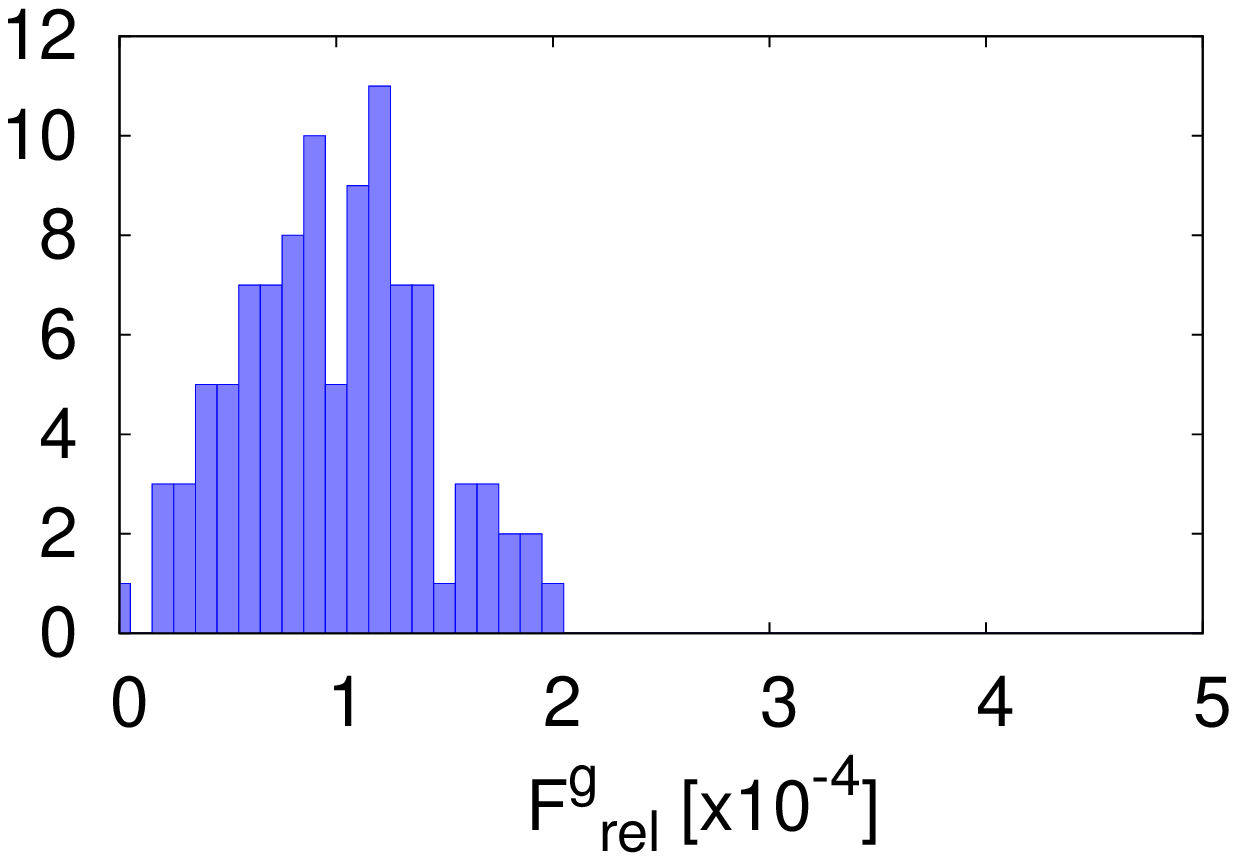}
	\includegraphics[width=0.32\textwidth]{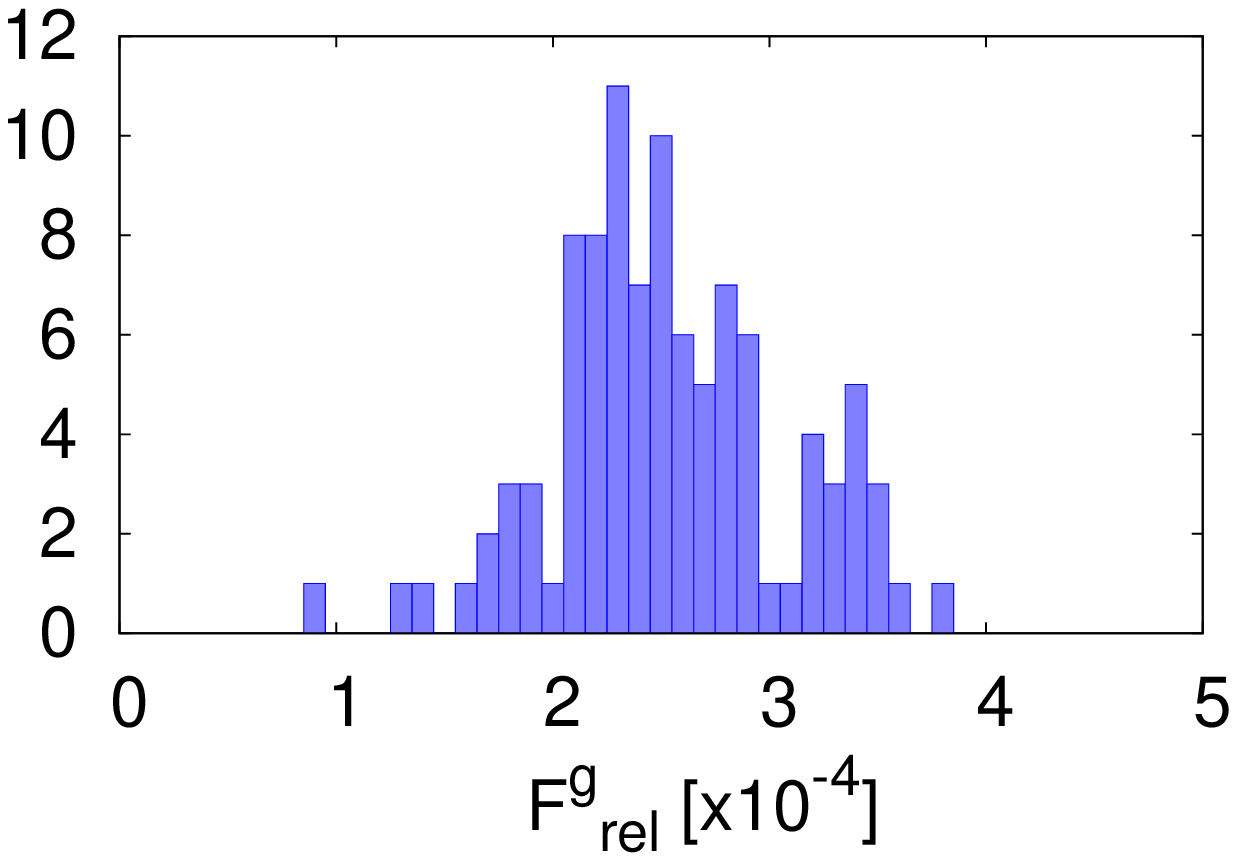}
	\includegraphics[width=0.32\textwidth]{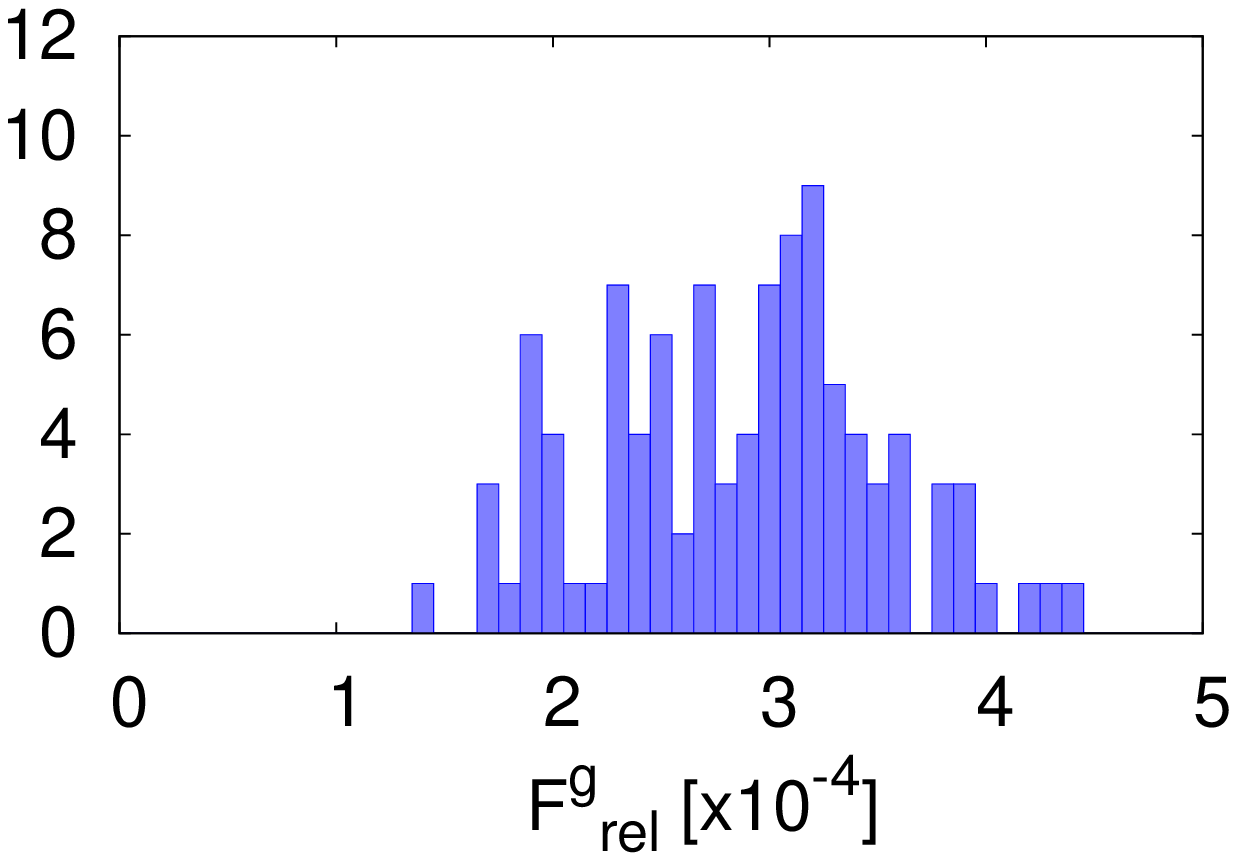}
	\caption{
		The relative deviation from the maximal gauge functional.
		From left to right with 10000, 3000 and zero simulated annealing steps.
		}
	\label{fig:SAlandau}
\end{figure}
We determined the maximum gauge functional value of all the runs, which we denote by 
$F^g_{\mathrm{max}}$ and define the relative deviation from it by
\be
	F^g_{\mathrm{rel}} = \frac{ F^g_{\mathrm{max}} - F^g}{F^g_{\mathrm{max}}}.
\ee
The latter is plotted in histograms in \fig{fig:SAlandau} for all the three runs.
The plot clearly demonstrates how the application of simulated annealing increases the chance
to find the global maximum, especially on a relatively large lattice of size $32^4$.
This test has been performed in parallel on two Tesla C2070 devices within several
hours.

\section{Summary}\label{sec:summary}
We presented a CUDA implementation for gauge fixing in lattice gauge field theories
based on the relaxation algorithms. 
The code is based on the cuLGT 
package\footnote{Both is available for downloaded under \href{http://www.culgt.com/}{www.cuLGT.com}.}
and supports the Landau, Coulomb and maximally Abelian gauge fixing conditions.

The implementation and the various optimization strategies have been discussed
in detail. We showed that simulated annealing and overrelaxation
can heavily be accelerated by employing GPUs.
We listed convergence results in different floating point precisions and
concluded that a mixed precision ansatz that performs only the critical
parts of the simulation in double precision is a good compromise 
in terms of precision ($\sim 10^{-5}$ relative to DP) and performance
($80\%-90\%$ of SP).

A maximum sustained performance of 370 GFlops on a single GTX~580 has been
reached and moreover linear scaling on 16 Tesla cards with 3.5 Teraflops, 
given that the number of time-slices per device does not fall below 16.

Lastly, we demonstrated how the combination of simulated annealing and the various
relaxation flavors can be tuned in such a way that either fast convergence to the gauge of choice
is reached or alternatively that a gauge functional value as high as possible is obtained.

We are currently preparing tests on the Kepler architecture, updates on Kepler 
performance will be available on our homepage shortly.

\section*{Acknowledgments}
We thank Giuseppe Burgio and Markus Quandt for helpful discussions
and we are very grateful to Gundolf Haase and Manfred Liebmann
for support with the ``mephisto'' cluster at the University of Graz.
M.S. is supported by the Research Executive 
Agency (REA) of the European Union under Grant Agreement 
PITN-GA-2009-238353 (ITN STRONGnet).

\appendix

\section{Counting flops}\label{sec:countingFlops}

As we discussed in \sec{algo_sumup}, the main work of Alg. \ref{alg1}
consists of applying the new update $g(x)$ to the neighboring links
of site $x$, i.e., Step 2 of the algorithm. 
We will now analyze this more quantitatively.
In \fig{fig:subgroupmult} we show the code snippet of cuLGT
for the multiplication of a $\SU{3}$ matrix with a $\SU{2}$ subgroup
element from the left.
Here, the $\SU{2}$ subgroup element is stored as a an object of
class \emph{Quaternion} (Cayley--Klein four parameter representation).

\begin{figure}[hb]
\begin{verbatimtab}
template<class Type> 
void SU3<Type>::leftSubgroupMult( lat_group_dim_t i, 
		lat_group_dim_t j, Quaternion<Real> *q )
{
	for( lat_group_dim_t k = 0; k < 3; k++ )
	{
		Complex<Real> IK = q->get( 0, 0 ) * get(i,k);
		IK += q->get( 0, 1 ) * get(j,k);

		Complex<Real> JK = q->get( 1, 0 ) * get(i,k);
		JK += q->get(1,1) * get(j,k);

		set( i, k , IK );
		set( j, k,  JK );
	}
}
\end{verbatimtab}
\caption{Multiplication of a $\SU{3}$ matrix by a $\SU{2}$ subgroup element in 
	Quaternion representation from the left.
	The total number of flop is 84 per $\SU{2}$ subgroup iteration;
	see discussion in the text.}
\label{fig:subgroupmult}
\end{figure}

As we can read of from the figure, in the loop over $k$ we encounter four 
complex multiplications (six flop each) 
plus two complex additions (two flop each), thus $28\cdot 3=84$ flop for
the update of $U_\mu(x)$ and equivalently for $U_\mu(x-\hat\mu)$.
Consequently, the number of flop for Step 2, in four dimensions,
sums up to $84\cdot2\cdot4=672$ per lattice site and $\SU{2}$ subgroup 
and hence to $672\cdot3=2016$ for $\SU{3}$.

As mentioned before, the above part is the same for all gauges and all update types.
Only Step 1 of Alg. \ref{alg1} distinguishes between different gauges and update types.
Let us consider for example an overrelaxation update for Landau gauge.
The latter consists of calculating $g(x)$ according to \eq{Kx} plus a first order
approximation of the exponentiation $g(x)\to g^\omega(x)$.
In the cuLGT code, the sum of \eq{Kx} is done on the \emph{Quaternion}
objects. Extracting the four reals of \emph{Quaternion} representation of 
a $\SU{2}$ subgroup element of $\SU{3}$ requires four flop, see \fig{fig:getSubgroupQuaternion}.
The \emph{Quaternion} objects are then gathered in an array in shared memory (\emph{shA})
according to \eq{Kx}. This means four flop (four additions) for each \emph{Quaternion}.
Thus the number of flop in \eq{Kx} is eight per link variable and in 4D eight link 
variables are involved, i.e., 64 flop per lattice site and $\SU{2}$ subgroup iteration
or 192 for $\SU{3}$.

\begin{figure}
\begin{verbatimtab}
template<class Type> 
Quaternion<Real> SU3<Type>::getSubgroupQuaternion( 
                 lat_group_dim_t iSub, lat_group_dim_t jSub )
{
	Quaternion<Real> q;
	Complex<Real> temp;
	temp = mat.get(iSub,iSub);
	q[0] = temp.x;
	q[3] = temp.y;
	temp = mat.get(jSub,jSub);
	q[0] += temp.x;
	q[3] -= temp.y;
	temp = mat.get(iSub,jSub);
	q[2] = temp.x;
	q[1] = temp.y;
	temp = mat.get(jSub,iSub);
	q[2] -= temp.x;
	q[1] += temp.y;

	return q;
}
\end{verbatimtab}
\caption{Extracting a $\SU{2}$ subgroup element of $\SU{3}$ in \emph{Quaternion} representation.}
\label{fig:getSubgroupQuaternion}
\end{figure}

Subsequently, the overrelaxation update  $g(x)\to g^\omega(x)$
is calculated.
Counting each operation in \fig{fig:orupdate} as one floating point operation (\emph{rsqrt} 
corresponds to two operations),
the \emph{effective} number of flop for the overrelaxation update is 22 per
lattice site and $\SU{2}$ subgroup, thus 66 for $\SU{3}$.

\begin{figure}
\begin{verbatimtab}
void OrUpdate::calculateUpdate( volatile Real (&shA)[4*NSB], 
	                                         short id )
{
	Real ai_sq = shA[id+NSB]   * shA[id+NSB]
	            +shA[id+2*NSB] * shA[id+2*NSB]
	            +shA[id+3*NSB] * shA[id+3*NSB];

	Real a0_sq = shA[id] * shA[id];

	Real b = (orParameter*a0_sq + ai_sq)/(a0_sq + ai_sq);
	Real c = rsqrt( a0_sq + b*b*ai_sq );

	shA[id] *= c;
	shA[id+NSB]   *= b*c;
	shA[id+2*NSB] *= b*c;
	shA[id+3*NSB] *= b*c;
}
\end{verbatimtab}
\caption{
	The overrelaxation update requires 22 flop per lattice site and $\SU{2}$ subgroup.
\label{fig:calculateUpdate}
}
\label{fig:orupdate}
\end{figure}

Summing up, the overrelaxation algorithm in $\SU{3}$ for Landau gauge requires
\begin{itemize}
	\item 192 flop to gather the neighboring links $U_\mu(x),\;U_\mu(x-\hat\mu)$,
	\item 66 flop for the overrelaxation update,
	\item 2016 flop to apply the new $g(x)$ to $U_\mu(x),\;U_\mu(x-\hat\mu)$
\end{itemize}
and thus in total 2274 flop/site. Note that we do not take the extra Flops for the reconstruction
of the third row of the $\SU{3}$ matrices into account.

For the heatbath kernel of the simulated annealing algorithm the number of flops cannot be calculated correctly because of the non-deterministic loops with random-number-dependent termination conditions. We counted the flops as if every loop is only run once and  each RNG call is counted as one flop. Both choices are very conservative. Therefore, a comparison of simulated annealing implementations should be based on pure time measurements.




\begin{thebibliography}{10}
\expandafter\ifx\csname url\endcsname\relax
  \def\url#1{\texttt{#1}}\fi
\expandafter\ifx\csname urlprefix\endcsname\relax\def\urlprefix{URL }\fi
\expandafter\ifx\csname href\endcsname\relax
  \def\href#1#2{#2} \def\path#1{#1}\fi

\bibitem{Neuberger:1986xz}
H.~Neuberger, {Nonperturbative Brs Invariance And The Gribov Problem},
  Phys.Lett. B183 (1987) 337.
\newblock \href {http://dx.doi.org/10.1016/0370-2693(87)90974-9}
  {\path{doi:10.1016/0370-2693(87)90974-9}}.

\bibitem{vonSmekal:2007ns}
L.~von Smekal, D.~Mehta, A.~Sternbeck, A.~G. Williams, {Modified Lattice Landau
  Gauge}, PoS LATTICE2007 (2007) 382.
\newblock \href {http://arxiv.org/abs/0710.2410} {\path{arXiv:0710.2410}}.

\bibitem{vonSmekal:2008es}
L.~von Smekal, A.~Jorkowski, D.~Mehta, A.~Sternbeck, {Lattice Landau gauge via
  Stereographic Projection}, PoS CONFINEMENT8 (2008) 048.
\newblock \href {http://arxiv.org/abs/0812.2992} {\path{arXiv:0812.2992}}.

\bibitem{Catterall:2011aa}
S.~Catterall, R.~Galvez, A.~Joseph, D.~Mehta, {On the sign problem in 2D
  lattice super Yang-Mills}, JHEP 1201 (2012) 108.
\newblock \href {http://arxiv.org/abs/1112.3588} {\path{arXiv:1112.3588}},
  \href {http://dx.doi.org/10.1007/JHEP01(2012)108}
  {\path{doi:10.1007/JHEP01(2012)108}}.

\bibitem{Mehta:2011ud}
D.~Mehta, S.~Catterall, R.~Galvez, A.~Joseph, {Supersymmetric gauge theories on
  the lattice: Pfaffian phases and the Neuberger 0/0 problem}, PoS LATTICE2011
  (2011) 078.
\newblock \href {http://arxiv.org/abs/1112.5413} {\path{arXiv:1112.5413}}.

\bibitem{Zwanziger:1998ez}
D.~Zwanziger, {Renormalization in the Coulomb gauge and order parameter for
  confinement in QCD}, Nucl.Phys. B518 (1998) 237--272.
\newblock \href {http://dx.doi.org/10.1016/S0550-3213(98)00031-5}
  {\path{doi:10.1016/S0550-3213(98)00031-5}}.

\bibitem{kirkpatrick}
S.~Kirkpatrick, C.~D. Gelatt~Jr., M.~P. Vecchi, {Optimization by Simulated
  Annealing}, Science 330 (1983) 671.

\bibitem{Bali:1996dm}
G.~Bali, V.~Bornyakov, M.~M{\"u}ller-Preussker, K.~Schilling, {Dual
  superconductor scenario of confinement: A Systematic study of Gribov copy
  effects}, Phys.Rev. D54 (1996) 2863--2875.
\newblock \href {http://arxiv.org/abs/hep-lat/9603012}
  {\path{arXiv:hep-lat/9603012}}, \href
  {http://dx.doi.org/10.1103/PhysRevD.54.2863}
  {\path{doi:10.1103/PhysRevD.54.2863}}.

\bibitem{Mehta:2010pe}
D.~Mehta, M.~Kastner, {Stationary point analysis of the one-dimensional lattice
  Landau gauge fixing functional, aka random phase XY Hamiltonian}, Annals
  Phys. 326 (2011) 1425--1440.
\newblock \href {http://arxiv.org/abs/1010.5335} {\path{arXiv:1010.5335}},
  \href {http://dx.doi.org/10.1016/j.aop.2010.12.016}
  {\path{doi:10.1016/j.aop.2010.12.016}}.

\bibitem{Mehta:2009zv}
D.~Mehta, A.~Sternbeck, L.~von Smekal, A.~G. Williams, {Lattice Landau Gauge
  and Algebraic Geometry}, PoS QCD-TNT09 (2009) 025.
\newblock \href {http://arxiv.org/abs/0912.0450} {\path{arXiv:0912.0450}}.

\bibitem{Hughes:2012hg}
C.~Hughes, D.~Mehta, J.-I. Skullerud, {Enumerating Gribov copies on the
  lattice}, Annals Phys. 331 (2013) 188--215.
\newblock \href {http://arxiv.org/abs/1203.4847} {\path{arXiv:1203.4847}}.

\bibitem{Mehta:2011xs}
D.~Mehta, {Finding All the Stationary Points of a Potential Energy Landscape
  via Numerical Polynomial Homotopy Continuation Method}, Phys.Rev. E84 (2011)
  025702.
\newblock \href {http://arxiv.org/abs/1104.5497} {\path{arXiv:1104.5497}},
  \href {http://dx.doi.org/10.1103/PhysRevE.84.025702}
  {\path{doi:10.1103/PhysRevE.84.025702}}.

\bibitem{Mehta:2011wj}
D.~Mehta, {Numerical Polynomial Homotopy Continuation Method and String Vacua},
  Adv.High Energy Phys. 2011 (2011) 263937.
\newblock \href {http://arxiv.org/abs/1108.1201} {\path{arXiv:1108.1201}},
  \href {http://dx.doi.org/10.1155/2011/263937}
  {\path{doi:10.1155/2011/263937}}.

\bibitem{Mehta:2012wk}
D.~Mehta, Y.-H. He, J.~D. Hauenstein, {Numerical Algebraic Geometry: A New
  Perspective on String and Gauge Theories}, JHEP 1207 (2012) 018.
\newblock \href {http://arxiv.org/abs/1203.4235} {\path{arXiv:1203.4235}},
  \href {http://dx.doi.org/10.1007/JHEP07(2012)018}
  {\path{doi:10.1007/JHEP07(2012)018}}.

\bibitem{Mandula:1987rh}
J.~Mandula, M.~Ogilvie, {The Gluon Is Massive: A Lattice Calculation of the
  Gluon Propagator in the Landau Gauge}, Phys.Lett. B185 (1987) 127--132.
\newblock \href {http://dx.doi.org/10.1016/0370-2693(87)91541-3}
  {\path{doi:10.1016/0370-2693(87)91541-3}}.

\bibitem{Mandula:1990vs}
J.~E. Mandula, M.~Ogilvie, {Efficient gauge fixing via overrelaxation},
  Phys.Lett. B248 (1990) 156--158.
\newblock \href {http://dx.doi.org/10.1016/0370-2693(90)90031-Z}
  {\path{doi:10.1016/0370-2693(90)90031-Z}}.

\bibitem{deForcrand:1989im}
P.~de~Forcrand, {Multigrid techniques for quark propagator},
  Nucl.Phys.Proc.Suppl. 9 (1989) 516--520.
\newblock \href {http://dx.doi.org/10.1016/0920-5632(89)90153-9}
  {\path{doi:10.1016/0920-5632(89)90153-9}}.

\bibitem{Clark:2009wm}
M.~Clark, R.~Babich, K.~Barros, R.~Brower, C.~Rebbi, {Solving Lattice QCD
  systems of equations using mixed precision solvers on GPUs},
  Comput.Phys.Commun. 181 (2010) 1517--1528.
\newblock \href {http://arxiv.org/abs/0911.3191} {\path{arXiv:0911.3191}},
  \href {http://dx.doi.org/10.1016/j.cpc.2010.05.002}
  {\path{doi:10.1016/j.cpc.2010.05.002}}.

\bibitem{Babich:2010mu}
R.~Babich, M.~A. Clark, B.~Joo, {Parallelizing the QUDA Library for Multi-GPU
  Calculations in Lattice Quantum Chromodynamics}\href
  {http://arxiv.org/abs/1011.0024} {\path{arXiv:1011.0024}}.

\bibitem{Babich:2011np}
R.~Babich, M.~Clark, B.~Joo, G.~Shi, R.~Brower, et~al., {Scaling Lattice QCD
  beyond 100 GPUs}\href {http://arxiv.org/abs/1109.2935}
  {\path{arXiv:1109.2935}}.

\bibitem{Alexandru:2011ee}
A.~Alexandru, C.~Pelissier, B.~Gamari, F.~Lee, {Multi-mass solvers for lattice
  QCD on GPUs}, J.Comput.Phys. 231 (2012) 1866--1878.
\newblock \href {http://arxiv.org/abs/1103.5103} {\path{arXiv:1103.5103}},
  \href {http://dx.doi.org/10.1016/j.jcp.2011.11.003}
  {\path{doi:10.1016/j.jcp.2011.11.003}}.

\bibitem{Bonati:2011dv}
C.~Bonati, G.~Cossu, M.~D'Elia, P.~Incardona, {QCD simulations with staggered
  fermions on GPUs}, Comput.Phys.Commun. 183 (2012) 853--863.
\newblock \href {http://arxiv.org/abs/1106.5673} {\path{arXiv:1106.5673}},
  \href {http://dx.doi.org/10.1016/j.cpc.2011.12.011}
  {\path{doi:10.1016/j.cpc.2011.12.011}}.

\bibitem{Chakrabarty:2012rv}
A.~Chakrabarty, P.~Majumdar, {Hybrid Monte Carlo with Wilson Dirac operator on
  the Fermi GPU}\href {http://arxiv.org/abs/1207.2223}
  {\path{arXiv:1207.2223}}.

\bibitem{Winter:2012wy}
F.~Winter, {Gauge Field Generation on Large-Scale GPU-Enabled Systems}, PoS
  LATTICE2012 (2012) 185.
\newblock \href {http://arxiv.org/abs/1212.0785} {\path{arXiv:1212.0785}}.

\bibitem{Winter:2011np}
F.~Winter, {Accelerating QDP++/Chroma on GPUs}\href
  {http://arxiv.org/abs/1111.5596} {\path{arXiv:1111.5596}}.

\bibitem{Schrock:2011hq}
M.~Schr{\"o}ck, {The chirally improved quark propagator and restoration of
  chiral symmetry}, Phys.Lett. B711 (2012) 217--224.
\newblock \href {http://arxiv.org/abs/1112.5107} {\path{arXiv:1112.5107}},
  \href {http://dx.doi.org/10.1016/j.physletb.2012.04.008}
  {\path{doi:10.1016/j.physletb.2012.04.008}}.

\bibitem{Schrock:2012rm}
M.~Schr{\"o}ck, H.~Vogt, {Gauge fixing using overrelaxation and simulated
  annealing on GPUs}, PoS LATTICE2012 (2012) 187.
\newblock \href {http://arxiv.org/abs/1209.4008} {\path{arXiv:1209.4008}}.

\bibitem{Cardoso:2012pv}
N.~Cardoso, P.~J. Silva, P.~Bicudo, O.~Oliveira, {Landau Gauge Fixing on GPUs},
  Comput.Phys.Commun. 184 (2013) 124--129.
\newblock \href {http://arxiv.org/abs/1206.0675} {\path{arXiv:1206.0675}},
  \href {http://dx.doi.org/10.1016/j.cpc.2012.09.007}
  {\path{doi:10.1016/j.cpc.2012.09.007}}.

\bibitem{Giusti:2001xf}
L.~Giusti, M.~Paciello, C.~Parrinello, S.~Petrarca, B.~Taglienti, {Problems on
  lattice gauge fixing}, Int.J.Mod.Phys. A16 (2001) 3487--3534.
\newblock \href {http://arxiv.org/abs/hep-lat/0104012}
  {\path{arXiv:hep-lat/0104012}}, \href
  {http://dx.doi.org/10.1142/S0217751X01004281}
  {\path{doi:10.1142/S0217751X01004281}}.

\bibitem{Stack:2002sy}
J.~Stack, W.~Tucker, R.~Wensley, {The maximal abelian gauge, monopoles, and
  vortices in SU(3) lattice gauge theory}, Nucl.Phys. B639 (2002) 203--222.
\newblock \href {http://dx.doi.org/10.1016/S0550-3213(02)00537-0}
  {\path{doi:10.1016/S0550-3213(02)00537-0}}.

\bibitem{CabibboMarinari1982}
N.~Cabibbo, E.~Marinari, {A New Method for Updating SU(N) Matrices in Computer
  Simulations of Gauge Theories}, Phys. Lett. B 119 (1982) 387.
\newblock \href {http://dx.doi.org/10.1016/0370-2693(82)90696-7}
  {\path{doi:10.1016/0370-2693(82)90696-7}}.

\bibitem{Salmon:2011:PRN:2063384.2063405}
J.~K. Salmon, M.~A. Moraes, R.~O. Dror, D.~E. Shaw, Parallel random numbers: as
  easy as 1, 2, 3, in: Proceedings of 2011 International Conference for High
  Performance Computing, Networking, Storage and Analysis, SC '11, ACM, New
  York, NY, USA, 2011, pp. 16:1--16:12.
\newblock \href {http://dx.doi.org/10.1145/2063384.2063405}
  {\path{doi:10.1145/2063384.2063405}}.

\bibitem{DeForcrand:1986af}
P.~De~Forcrand, D.~Lellouch, C.~Roiesnel, {OPTIMIZING A LATTICE QCD SIMULATION
  PROGRAM}, J.Comput.Phys. 59 (1985) 324--330.
\newblock \href {http://dx.doi.org/10.1016/0021-9991(85)90149-4}
  {\path{doi:10.1016/0021-9991(85)90149-4}}.

\bibitem{FermiQCD}
M.~Di~Pierro, et~al., {www.fermiqcd.net}, Nucl. Phys. Proc. Suppl. 129 (2004)
  832--834.
\newblock \href {http://arxiv.org/abs/hep-lat/0311027}
  {\path{arXiv:hep-lat/0311027}}.

\end{thebibliography}



\end{document}